\documentclass[journal=aaemcq,manuscript=article]{achemso}
\usepackage[version=3]{mhchem} 
\usepackage{orcidlink}
\usepackage{hyperref}

\usepackage{amsmath, amssymb}
\usepackage{cleveref}

\usepackage{tabularx} 
\usepackage{graphicx} 
\usepackage{siunitx}

\hypersetup{
  colorlinks=false,      
  linkcolor=black,      
  citecolor=black,
  urlcolor=black
}

\crefname{equation}{Equation}{Equations}
\Crefname{equation}{Equation}{Equations}

\definecolor{figcolor}{RGB}{0,102,204}   
\definecolor{tabcolor}{RGB}{0,153,0}     
\definecolor{algcolor}{RGB}{204,0,102}   %

\usepackage{array}
\usepackage{booktabs}
\usepackage{tabularx}
\usepackage{makecell}

\newcolumntype{Y}{>{\centering\arraybackslash}X}

\newcolumntype{C}[1]{>{\centering\arraybackslash}m{#1}}

\usepackage[svgnames]{xcolor}

\makeatletter
\newenvironment{datastatement}{%
\acs@section*{\datastatementname}%
}{}
\newcommand*\datastatementname{Data and Code Availability Statement}
\makeatother

\makeatletter
\newenvironment{conflictstatement}{%
\acs@section*{\conflictstatementname}%
}{}
\newcommand*\conflictstatementname{Notes}
\makeatother

\SectionNumbersOff

\author{Arpan Sur\,\orcidlink{0009-0003-5833-9998}}

\author{Kawshik Nath\,\orcidlink{0009-0001-9749-278X
}}
\altaffiliation{Department of Electrical and Electronic Engineering, Chittagong University of Engineering and Technology, Chittagong, Chittagong-4349, Bangladesh}

\author{Ahmed Zubair\,\orcidlink{0000-0002-1833-2244}}
\email{ahmedzubair@eee.buet.ac.bd}
\affiliation[BUET]
{Department of Electrical and Electronic Engineering, Bangladesh University of Engineering and Technology, Dhaka, Dhaka-1205, Bangladesh}

\title{Nature-Inspired Hyperuniform Nanohole Patterning for Robust Broadband Absorption Enhancement in Perovskite Solar Cells}


\begin{document}

\begin{tocentry}

\includegraphics[width=3.25in,height=1.75in]{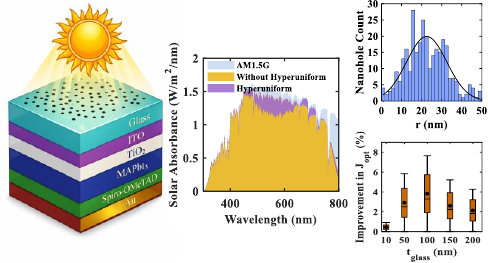}





\end{tocentry}


\begin{abstract}
Nature-inspired hyperuniform disorder offers a promising route to broadband light trapping in ultrathin perovskite solar cells by avoiding narrowband, illumination-sensitive responses commonly associated with periodic nanophotonic textures. Here, we introduce a nature-inspired ingenious hyperuniform nanohole architecture integrated into the front glass of a planar MAPbI$_3$ perovskite solar cell, serving as a junction-preserving strategy to enhance optical absorption and photovoltaic performance. In comparison with planar and periodic textures, the hyperuniform architecture redistributed incident light across a broader spectrum of in-plane momentum states, strengthened near-interface electromagnetic fields, and improved long-wavelength coupling into the absorber, thereby increasing the effective optical path length without altering the electronically active interfaces. To quantify these effects, we employed a coupled three-dimensional multiphysics framework that integrates finite-difference time-domain (FDTD) optical simulations with drift-diffusion electrical modeling. The optimized design exhibited broadband absorption enhancement, weak polarization dependence, and strong angular tolerance, while suppressing interference-driven spectral oscillations and reducing sensitivity to patterned-layer thickness. Relative to the planar structure, the hyperuniform architecture increased the short-circuit current density from 21.57 to 23.92 mAcm$^{-2}$ and improved the power conversion efficiency from 21.03\% to 23.62\%, while maintaining $\mathrm{V_{oc}}$ at 1.13 V and preserving a high fill factor of 87.66\%. In addition to statistical pattern-invariant performance, stochastic radius-variation analysis indicated a positive enhancement in photocurrent and under fabrication-relevant dimensional disorder. These results establish front-substrate hyperuniform nanohole patterning as a robust and practical light-management strategy for high-performance perovskite photovoltaics.
\end{abstract}

\section{Introduction}
Perovskite semiconductors have become a burgeoning research topic in emerging optoelectronic technologies~\cite{Zhang2023NML} as these materials enable a paradigm shift in the development of high-efficiency photovoltaics~\cite{NREL_CellEfficiency, rahman2025arxiv, nath2026high}, next-generation light-emitting diodes~\cite{zheng2024ultralow}, and ultra-sensitive photodetectors~\cite{tsarev2025vertically} through low-cost solution processing. 
They exhibit higher optical absorption, lower recombination rates, longer carrier diffusion lengths, and greater defect tolerance compared to other emerging PV platforms.
Moreover, they offer a tunable bandgap, enabling spectral matching and flexible device integration, thereby improving their utility in flexible and semi-transparent applications~\cite{han2025perovskite,Shilpa2023,Afre2024}.
Due to these benefits, perovskite solar cells (PSCs) have enabled unprecedented global research engagement and accelerated device development at a pace that surpasses that of traditional photovoltaic technologies.
These efforts led to a jump in power conversion efficiency (PCE) of PSCs from 3.8\% to 25.5\% over the past 16 years, approaching the theoretical Shockley-Queisser limit of around 30\% \cite{Kojima2009,Michael,NREL_CellEfficiency}. 
Among the many perovskite materials explored to date, methylammonium lead iodide (CH$_3$NH$_3$PbI$_3$), or MAPbI$_3$, is the most widely studied absorber material. 
MAPbI$_3$ has a direct bandgap ($\mathrm{E_g}$) of 1.5--1.6 eV, a high absorption coefficient ($\geq 10^{4}~\text{cm}^{-1}$), a substantial dielectric constant, and excellent solubility for processing. However, the presence of toxic Pb in MAPbI$_3$ strongly discourages the use of a thicker MAPbI$_3$ layer in solar cells. 
Fortunately, the high absorption coefficient of MAPbI$_3$ enables a sub-300-nm absorber layer to efficiently absorb light up to 600 nm.
However, wavelengths longer than 600 nm are often absorbed less efficiently than shorter ones. Therefore, additional light-management systems are required to achieve efficient broadband absorption.


Efficient light management remains a critical challenge in ultrathin absorber layers because of reduced optical path lengths. Various nanophotonic strategies have been investigated to address these limitations, including gratings \cite{AHMADI2025108333, nakti2023ultra}, nanostructuring \cite{haque2022photonic,rahman2026hierarchically}, plasmonic nanoparticles \cite{Luo2017}, and spectral up- or down-conversion layers \cite{haque2024photon,akhtary2023high}. 
Mohammadi \textit{et al.} designed an Ag nanostructure-based MAPbI$_3$ PSC, reporting PCE of 16.20\% \cite{Mohammadi2023}. 
To enhance light trapping and carrier collection, Chen \textit{et al.} proposed inverse-opal nanostructuring in MAPbI$_3$ PSC, increasing the PCE by 12.12\% \cite{Chen}. 
Moreover, Wang \textit{et al.} integrated nanoimprinting with a dual-functional, nanostructured PEAI-based 2D/3D interlayer, achieving a PCE of 22.45\% \cite{WANG2024155686}.
Rahman \textit{et al.} proposed a theoretically optimized MAPbI$_3$ double-hole-layer PSC featuring CuO/I$_2$O$_5$-doped Spiro-OMeTAD and hierarchical ellipsoidal texturing to enhance band alignment, hole extraction, and optical confinement \cite{rahman2026hierarchically}. Their coupled non-isothermal opto-electro-thermal analysis predicted a PCE of 24.93\% for the thermally operated device, resulting in a 15.15\% improvement in PCE, highlighting both high performance and thermal robustness.
Although these approaches can deliver significant optical gains, they often rely on multi-step nanofabrication processes or complex materials integration, which complicates their translation to low-cost, high-throughput, and large-area manufacturing. 

Nanohole patterning has emerged as a relatively scalable alternative \cite{HE2013265}; however, conventional implementations have notable drawbacks.  
Periodic nanohole lattices typically rely on angle- and wavelength-selective resonances, making performance sensitive to illumination conditions. 
In contrast, fully random patterns may introduce uncontrolled clustering and substantial device-to-device variability, thereby compromising reproducibility and predictable optical response. 

Hyperuniform disordered structures provide an intermediate and potentially superior design strategy. These systems are statistically disordered yet exhibit a form of ``hidden order" characterized by suppressed long-wavelength density fluctuations \cite{Florescu}. Such control enables engineering of the structure factor so that optical momentum is distributed broadly and nearly isotropically in reciprocal space. Importantly, this short-range geometric order has been demonstrated to support complete photonic band gaps in disordered dielectric networks, thereby inhibiting light propagation in all directions and polarizations. This behavior resembles that of photonic crystals but does not require long-range translational periodicity. Inspired by naturally occurring photonic architectures observed in biological systems—such as the quasi-hyperuniform arrangement of photoreceptors in chicken's retina \cite{Jiao} and the looped vein networks of tree leaves \cite{Liu}, these designs have been extended to various photonic applications, including waveguides, metalenses, polarizers, and topological insulators \cite{Islam, Zhou, Mitchell}. By employing this concept, Jeronimo \textit{et al.} achieved a remarkable PCE of 22.35\% in an ultrathin GaAs solar cell utilizing quasi-random photonic-crystal patterning that strengthened broadband optical absorption \cite{BUENCUERPO2022107080}. Milena \textit{et al.} further demonstrated that stealthy hyperuniform surface textures in organic solar cells can simultaneously enhance light absorption and device performance \cite{Milena}. Recently, Taakoli \textit{et al.} experimentally demonstrated a significant enhancement of solar absorption in a Si solar cell by incorporating a hyperuniform architecture \cite{Tavakoli}. 
However, their integration into perovskite-based photovoltaics still remains uncharted.

In this work, we addressed this gap by introducing a hyperuniform nanohole texture on the top surface of a planar MAPbI$_3$ perovskite solar cell as a junction-preserving light-management strategy. 
At first, a hyperuniform nanohole topology was integrated into the front glass of a planar MAPbI$_3$ perovskite solar cell to enable enhanced optical coupling while leaving the charge-selective interfaces unmodified. 
To quantify its impact, we combined three-dimensional finite-difference time-domain (FDTD) simulations with three-dimensional drift--diffusion modeling to calculate absorbed power density, spatial photogeneration, near-field redistribution, and the resulting photovoltaic figures of merit, including $\mathrm{J_{sc}}$, $\mathrm{V_{oc}}$, FF, and PCE. 
This multiphysics framework further enabled us to elucidate the light-trapping physics of the hyperuniform architecture, particularly its ability to enhance long-wavelength coupling, strengthen near-interface fields, and suppress interference-driven spectral oscillations. 
In addition, we evaluated the optical response under varying polarization and incident angle to assess polarization insensitivity and angular robustness, and performed systematic sweeps of nanohole radius and patterned-layer height to identify optimum geometric features. 
We also benchmarked the design against planar and periodic nanohole references. 
Finally, we evaluated robustness against fabrication-related dimensional disorder by analyzing equivalent hyperuniform nanohole patterns and applying stochastic radius perturbations to them. 
Overall, this study establishes hyperuniform front-substrate nanohole patterning as a practical route to broadband absorption enhancement and performance stability in perovskite photovoltaics.

\section{Design and Methodology}
\subsection{Device Geometry}

\begin{figure}[!htb]
\centering\includegraphics[width=\textwidth]{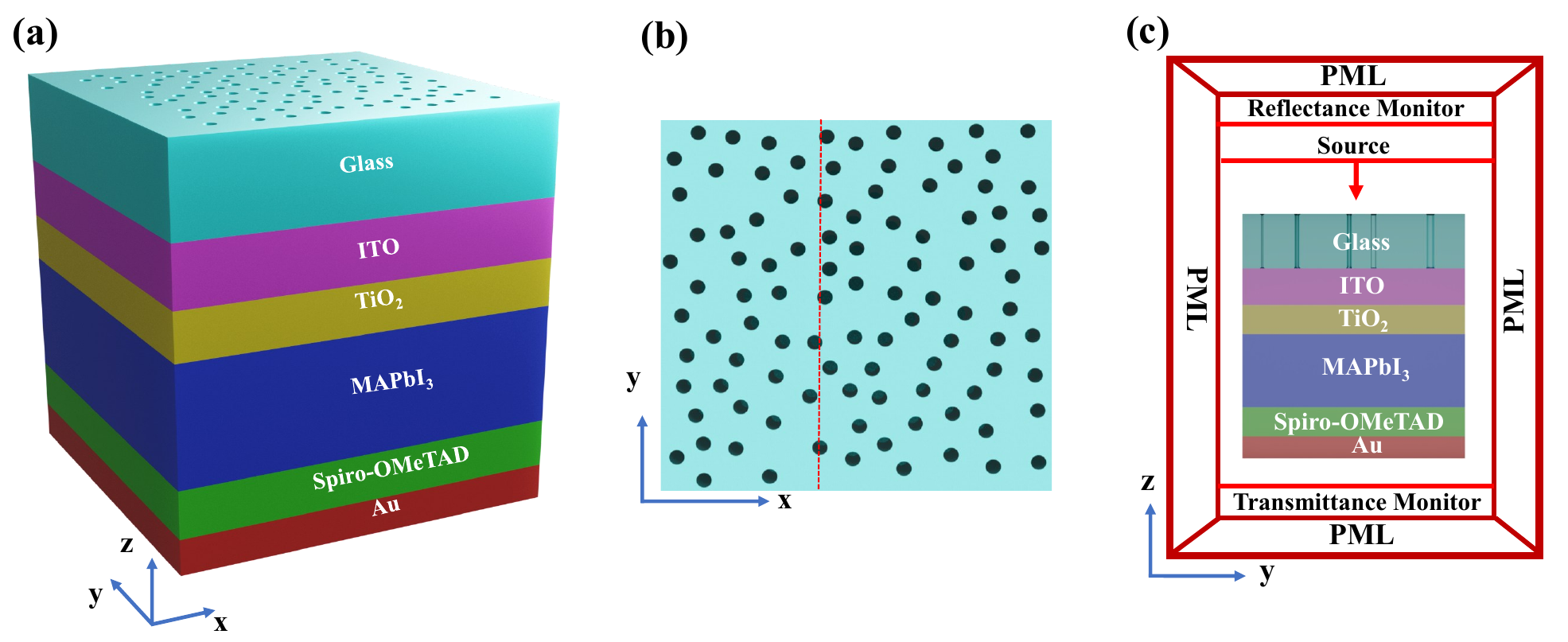}
\caption{Proposed hyperuniform-patterned MAPbI$_3$ perovskite solar cell and the optical simulation domain. (a) Three-dimensional schematic of the device stack, consisting of glass/ITO/TiO$_2$/MAPbI$_3$/Spiro-OMeTAD/Au, with the hyperuniform nanohole texture patterned into the incident-side glass layer. (b) Top view of the hyperuniform nanohole distribution in the x-y plane; the red line indicates the section used for the cross-sectional view. (c) Cross-sectional view of the FDTD simulation region and proposed device in the y-z plane taken along the line marked in panel (b). The simulation region includes broadband source, reflectance, and transmittance monitors, as well as perfectly matched layer (PML) boundaries.}
\label{fig:structure}
\end{figure}


\autoref{fig:structure}(a) illustrates the designed planar heterojunction perovskite solar cell architecture, comprising Glass/ITO/TiO$_2$/MAPbI$_3$/Spiro-OMeTAD/Au. 
To avoid the surface imperfections and defect formation that can accompany nanopatterning of electronically active interfaces and thereby degrade carrier transport, the hyperuniform nanohole texture was implemented in the incident-side glass of the planar solar cell rather than in the intermediate functional layers.
The MAPbI$_3$ layer was chosen as the primary photoactive absorber due to its direct bandgap of approximately 1.55~eV and large extinction coefficient in the visible range, which enables strong absorption across the visible spectrum while supporting high photovoltaic conversion efficiency \cite{eperon2014formamidinium}. 
Charge-selective carrier extraction was achieved by using TiO$_2$ as the electron-transport layer (ETL) and Spiro-OMeTAD as the hole-transport layer (HTL). 
With a wide bandgap of about 3.2~eV and a favorable conduction-band alignment with MAPbI$_3$, TiO$_2$ facilitates electron collection at the transparent ITO front electrode while suppressing hole transport \cite{Dong2024}. 
On the hole-extraction side, Spiro-OMeTAD was employed as the HTL to enhance hole selectivity and interfacial passivation. 
The device was completed with ITO and Au serving as the transparent front and reflective back electrodes, respectively.

The material interfaces in this stack are supported by prior experimental demonstrations reported in the literature, including TiO$_2$/MAPbI$_3$ and MAPbI$_3$/Spiro-OMeTAD contact schemes \citep{Dong2024, Galatopoulos2017, Vo2024, daoudi2024outcomes}, thereby confirming the structural feasibility of the baseline configuration used for subsequent optical and photovoltaic analyses. 
The complex refractive index profiles of all materials are collected from prior experimental studies \cite{Raoult2019, Konig2014, Phillips2015, Palik1998} and are further plotted in Figure S1 of Supplementary Information. 
The electrical properties of each layer in the simulation were taken from prior studies and described in detail in Table S1. The optimized thickness values of each layer of the planar structure were calculated by sweeping the layer thickness, as shown in Section S3 of the Supplementary Information. The optimized values are shown in \autoref{tab:solar_cell_layers}.

The hyperuniform nanohole centers were generated using a shuffled-lattice construction with bounded in-cell displacement, followed by enforcement of a minimum-separation constraint to prevent hole overlap \cite{bridson2007fast}.
This algorithm also guaranteed a minimum distance between neighboring nanoholes by creating a grid layout in which each cell contains a single nanohole, thereby supporting the suppression of longer wavelengths (lower wavevectors). 
The detailed algorithm is described in Supplementary Information.
This point pattern was used to create a hyperuniform nanohole layer in the FDTD simulation environment for further analysis. A top view and cross-sectional view of the generated nanohole-patterned solar cell are illustrated in \autoref{fig:structure}(b) and (c). The detailed fabrication process for this patterned surface is presented in a later section.

\begin{table}[!htb]
\caption{Layer dimensions and corresponding functions of the solar cell stack.}
\label{tab:solar_cell_layers}
\centering
\setlength{\tabcolsep}{5pt}        
\renewcommand{\arraystretch}{1.25}  
\begin{tabular}{%
>{\centering\arraybackslash}m{0.25\columnwidth}
>{\centering\arraybackslash}m{0.12\columnwidth}
>{\centering\arraybackslash}m{0.12\columnwidth}
>{\centering\arraybackslash}m{0.40\columnwidth}}
\toprule
Material & Symbol & Thickness & Functionality \\
\midrule
Glass        & $\mathrm{t_{glass}}$ & \SIrange{10}{200}{\nano\meter} & Substrate, hyperuniform patterning and anti-reflective coating \\
ITO          & $\mathrm{t_{tco}}$   & \SI{70}{\nano\meter}           & Transparent front contact \\
TiO$_2$      & $\mathrm{t_{etl}}$   & \SI{90}{\nano\meter}           & Electron transport layer \\
MAPbI$_3$    & $\mathrm{t_{psk}}$   & \SI{200}{\nano\meter}          & Primary absorber layer \\
Spiro-OMeTAD & $\mathrm{t_{htl}}$   & \SI{30}{\nano\meter}           & Hole transport layer \\
Au           & $\mathrm{t_{bc}}$    & \SI{150}{\nano\meter}          & Back contact \\
\bottomrule
\end{tabular}
\end{table}

\subsection{Simulation Methodology}

To evaluate the optical field response (${\vec{\mathrm{E}}_{\mathrm{op}}}$) of the proposed structure, Maxwell’s curl equations, Equations \ref{eq:max3}-\ref{eq:max4}, were solved using the three-dimensional FDTD method. 
All optical simulations were carried out in Ansys Lumerical FDTD to obtain the absorbed power density ($\mathrm{P_{abs}}$) and carrier generation rate ($\mathrm{G}$).
\vspace{-30pt}

    \begin{align}
    \label{eq:max3}
    &\frac{\partial\vec{\mathrm{D}}}{\partial \mathrm{t}}=\nabla\times\vec{\mathrm{H}},\\[4pt]
    \label{eq:max4}
    &\frac{\partial\vec{\mathrm{H}}}{\partial \mathrm{t}}=-\frac{1}{\mathrm{\mu_0}}\;\nabla\times\vec{\mathrm{E}}_{\mathrm{op}}.
    \end{align}
    
\vspace{-10pt}

Here, $\vec{\mathrm{H}}$, $\vec{\mathrm{D}}$, $\mu_0$, and t  are the magnetic field, electric displacement fields, vacuum permeability, and time, respectively. The displacement field is defined as $\vec{\mathrm{D}}=\varepsilon\vec{\mathrm{E}}_{\mathrm{op}}$, where, $\varepsilon$ is the complex relative dielectric constant. The optical response of each constituent layer was described by its refractive index ($n$) and extinction coefficient ($\kappa$), with $\varepsilon = (n+i\kappa)^2$. Using the calculated $\vec{\mathrm{E}}_{\mathrm{op}}(\vec{\mathrm{r}}, \omega)$ throughout the computational domain, the $\mathrm{P_{abs}}(\vec{\mathrm{r}}, \omega)$ and $\mathrm{G}(\vec{\mathrm{r}})$ were determined using Equations \ref{eq:Pabs}-\ref{eq:Gint}.
\vspace{-40pt}

\begin{gather}
    \label{eq:Pabs}
    \mathrm P_{\mathrm{abs}}=-\frac{1}{2}\,\omega\,\left| \vec{\mathrm E}_{\mathrm{op}}( \vec{\mathrm r},\omega)\right|^2
    \,\Im\{\epsilon(\vec{\mathrm r},\omega)\}, \\[4pt]
    \label{eq:g}
    \mathrm g(\vec{\mathrm r},\omega)=\frac{\mathrm{P}_{\mathrm{abs}}}{\mathrm{\hbar} \omega}
    =-\frac{\pi}{\mathrm{h}}\left|\vec{\mathrm E}_{\mathrm{op}}(\vec{\mathrm r},\omega)\right|^2
    \,\Im\{\epsilon(\vec{\mathrm r},\omega)\}, \\[4pt]
    \label{eq:Gint}
    \mathrm G(\vec{\mathrm r})=\int \mathrm g(\vec{\mathrm r},\omega)\,d\omega.
\end{gather}

 \vspace{-10pt}

Here, $\vec{\mathrm r}$, $\omega$, $\hbar$, and $\Im\{\epsilon\}$ are the position vector, angular frequency, reduced Planck constant, and 
imaginary part of the complex dielectric constant, respectively.  
Assuming unity internal quantum efficiency (each absorbed photon was converted into one electron-hole pair), we calculated the optical current density, $\mathrm{J_{opt}}$ using \autoref{eq:J_optical}:

\begin{equation}
    \mathrm {J_{opt}} = \mathrm{q}\int \frac{\lambda \mathrm{P_{abs} (\lambda) I_{AM1.5} (\lambda)}}{\mathrm{hc P_{in}(\lambda)}} d\lambda.
    \label{eq:J_optical}
\end{equation}

Here, $\mathrm{h}$, $\mathrm{c}$, $\lambda$, $\mathrm{I_{AM1.5}(\lambda)}$, $\mathrm{P_{in}(\lambda)}$, and $\mathrm{q}$ denote Planck's constant, the speed of light, light wavelength, AM1.5 solar spectral irradiance, incident optical power spectrum, and the electron charge, respectively. A y-z cross-section of the 3D FDTD model is illustrated in \autoref{fig:structure}(c). 
A $1\mu m \times 1\mu m$ solar cell stack was defined in the x-y direction for hyperuniform nanohole pattern definition.
A 200 nm distance was maintained between the structure, the monitor, and the simulation region to ensure consistent mesh spacing across the different objects in the simulation region.
The model employed perfectly matched layers (PMLs) in all directions to eliminate spurious reflections at the simulation boundaries.

A broadband plane-wave source spanning 300-800 nm excited the device, with the upper wavelength limit corresponding to the bandgap of the primary perovskite absorber (1.55 eV). 
A uniform mesh was used for improved accuracy, with grid sizes of 5 nm in the $\mathrm {x-y}$ plane and 1 nm along the $\mathrm{z}$-direction. 
Finally, we positioned frequency-domain power monitors above and below the device to record the reflectance, $\mathrm{R(\lambda)}$, and transmittance, $\mathrm{T(\lambda)}$, respectively, enabling the extraction of absorption and the corresponding photocurrent. 

Later, the calculated spatial photo-generation rate, $\mathrm G(\vec{\mathrm r})$, was used to compute the electrical solar cell characteristics: short-circuit current ($\mathrm{J_{sc}}$), open-circuit voltage ($\mathrm{V_{oc}}$), fill factor (FF), and PCE. 
To obtain these parameters, we solved Poisson's equation (\autoref{eq:poisson}), the continuity equations (Equations \ref{eq:drift_n}-\ref{eq:drift_p}), and the drift-diffusion equations (Equations \ref{eq:cont_n}-\ref{eq:cont_p}) self-consistently with the Ansys Lumerical Charge solver using the finite element method (FEM). 
Solving the equations yields the current density (J)-voltage (V) characteristics and the power density (P)-voltage (V) characteristics.

\vspace{-40pt}
\begin{align}
\label{eq:poisson}
&-\nabla \cdot (\varepsilon_{\mathrm{dc}} \nabla \mathrm V) = \mathrm q \rho, \\[4pt]
\label{eq:drift_n}
&\frac{\partial \mathrm n}{\partial \mathrm t} = \frac{1}{\mathrm q} \nabla\cdot \vec{\mathrm J}_{\mathrm n} - \mathrm {R_n + G_n}, \\[4pt]
\label{eq:drift_p}
&\frac{\partial \mathrm p}{\partial \mathrm t} = -\frac{1}{\mathrm q} \nabla\cdot \vec{\mathrm J}_\mathrm {p} - \mathrm {R_p + G_p}, \\[4pt]
\label{eq:cont_n}
&\vec{\mathrm J}_\mathrm{n} = \mu_\mathrm{n} \big(\mathrm q\vec{\mathrm E}_\mathrm{n} + \mathrm {\mathrm {k\nabla T} \big) }\mathrm n + \mu_\mathrm{n} \mathrm{kT} \nabla \mathrm n, \\[4pt]
\label{eq:cont_p}
&\vec{\mathrm J}_\mathrm{p} = \mu_\mathrm{p} \big(\mathrm q\vec{\mathrm E}_\mathrm{p} + \mathrm {\mathrm {k\nabla T} \big) }\mathrm p + \mu_\mathrm{p} \mathrm{kT} \nabla \mathrm p. 
\end{align}

\vspace{-15pt}

In Equations ~\ref{eq:poisson}--\ref{eq:cont_p}, $\rho$ is the space-charge density, $\mu_\mathrm{n}$ and $\mu_\mathrm{p}$ denote the electron and hole mobilities, $\mathrm{R_n}$ and $\mathrm{R_p}$ are the recombination rates, and $\mathrm{G_n}$ and $\mathrm{G_p}$ are the generation rates. The quantities $\vec{\mathrm{E}}_\mathrm{n}$ and $\vec{\mathrm{E}}_\mathrm{p}$ represent the electric fields experienced by electrons and holes, respectively, whereas $\vec{\mathrm{J}}_\mathrm{n}$ and $\vec{\mathrm{J}}_\mathrm{p}$ denote the corresponding current densities. Moreover, $\mathrm{k}$ is the Boltzmann constant, $\mathrm{T}$ is the absolute temperature, $\mathrm{V}$ is the electrostatic potential, and $\varepsilon_{\mathrm{dc}}$ is the static dielectric permittivity. The subscripts $\mathrm{n}$ and $\mathrm{p}$ refer to electron and hole quantities, respectively, while $\mathrm{n}$ and $\mathrm{p}$ denote the corresponding carrier concentrations.
To further analyze our structure, we measured its absorbance response at all intermediate polarization angles within TE- and TM-polarizations. 
Additionally, we calculated the absorption response at different incident angles to characterize this structure's light-trapping ability at oblique angles. To do that, we used the broadband fixed angle source technique (BFAST) boundary and source conditions to ensure the actual injection angle remains constant across our broad operating wavelength range.
We further varied the nanohole radius and thickness to assess the effect of these geometric variations on the absorbance of hyperuniform patterning.
Furthermore, to assess the impact of experimental variations in nanopatterning arising from fabrication imperfections, we added a Gaussian distribution to the hyperuniform nanohole radii.
For these analyses, the Python API of Lumerical FDTD was used to calculate the optical response of the solar cell. 







\section{Results and Discussion}

\subsection{Solar Absorption Enhancement through Hyperuniform Patterning}

\begin{figure}[!htb]
\centering\includegraphics[width=1\textwidth]{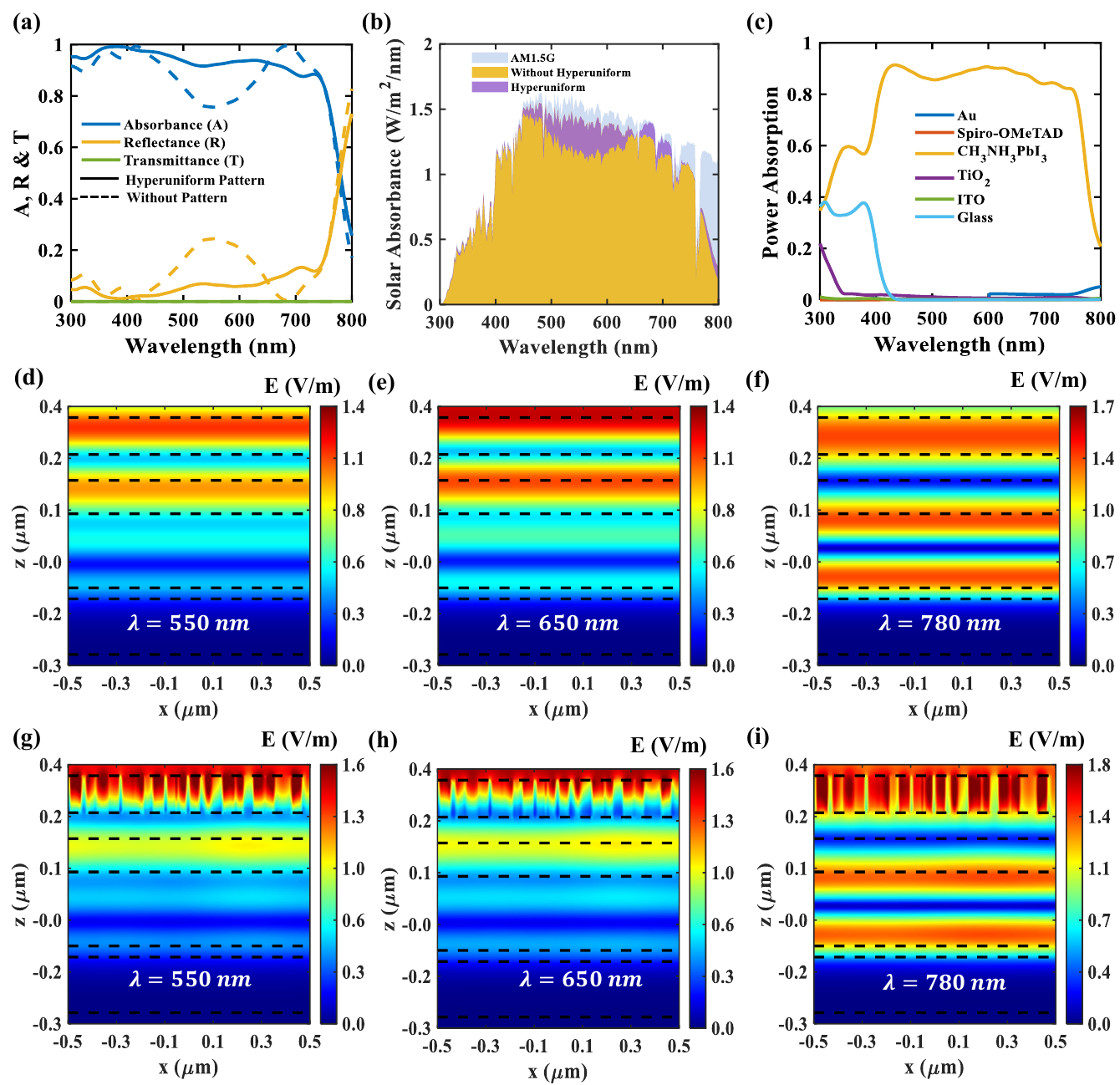}
\caption{Optical response and near-field distributions of the planar and hyperuniform-patterned MAPbI$_3$ perovskite solar cells. (a) Absorbance ($\mathrm A$), reflectance ($\mathrm R$), and transmittance ($\mathrm T$) spectra of the optimized hyperuniform-patterned device and its planar reference counterpart, respectively. (b) Solar-weighted absorbance obtained from the simulated absorbance response under the AM1.5G spectrum, comparing the planar and hyperuniform structures. (c) Normalized power absorption of different layers of a hyperuniform patterned solar cell. (d--f) Cross-sectional electric-field intensity distributions of the planar device at $\lambda = 550$, 650, and 780 nm, respectively. (g--i) Corresponding electric-field intensity distributions of the hyperuniform-patterned device at the same wavelengths. The hyperuniform front texture strengthens near-interface fields and promotes more effective long-wavelength light coupling into the absorber, consistent with enhanced broadband light trapping.}
\label{fig:field}
\end{figure}

\autoref{fig:field} compares the spectral optical response of the planar and hyperuniform-patterned devices and relates it to the underlying near-field distribution. 
As shown in \autoref{fig:field}(a), planar structures maintained strong absorption in the lower visible wavelength range, while reflectance increased at longer wavelengths, near the MAPbI$_3$ band edge. 
However, the hyperuniform texture redistributed the spectral response beneficially, yielding improved absorption in the red and near-band-edge region where single-pass absorption in a thin perovskite layer is inherently weaker. 
This behavior is clearly reflected in the solar-weighted absorbance comparison in \autoref{fig:field}(b), which shows that the hyperuniform architecture more effectively utilizes the AM1.5G photon flux across a wide range of the solar spectrum. \autoref{fig:field}(c) illustrates the normalized power absorption across different layers of the hyperuniform patterned solar cell. 
Lower wavelengths are trapped within the glass layer, while the absorption of longer wavelengths in the active layer significantly increases in the presence of hyperuniform patterning, thereby boosting the overall solar photocurrent yield.

The physical origin of this enhancement is evident from the electric-field intensity profiles in \autoref{fig:field}(d--i). For the planar structure [\autoref{fig:field}(d--f)], the field profile was mainly influenced by thin-film interference, leading to laterally uniform quasi-standing-wave patterns across the multilayer stack. At $\lambda = 550$ nm, the optical field was rapidly attenuated inside the absorber because MAPbI$_3$ already exhibited strong intrinsic absorption in this spectral range. As the wavelength increases to 650 and 780 nm, the planar device showed more pronounced interference fringes and weaker optical-field penetration into the active region. This phenomenon indicates limited light trapping near the band edge.

In contrast, the hyperuniform-patterned device [\autoref{fig:field}(g--i)] showed significant field localization at the textured glass interface, followed by increased penetration of the optical field into the underlying ITO/TiO$_2$/MAPbI$_3$ stack. 
The nanohole texture scattered the incident wave into a broader in-plane momenta distribution, which facilitates coupling into oblique and partially guided optical channels that are not efficiently accessed in the planar device. 
As a result, the field distribution became less constrained by a single interference condition and was more effectively redistributed throughout the absorber, particularly at 650 nm (red) and 780 nm (IR). 
These results indicated that hyperuniform patterning enhances active-layer absorption by improving optical coupling in the in-plane direction and increasing the effective optical path length without altering the charge-selective layers.

\subsection{Modulation of Carrier Generation by Hyperuniform Patterning}

\begin{figure}[!htb]
\centering\includegraphics[width=\textwidth]
{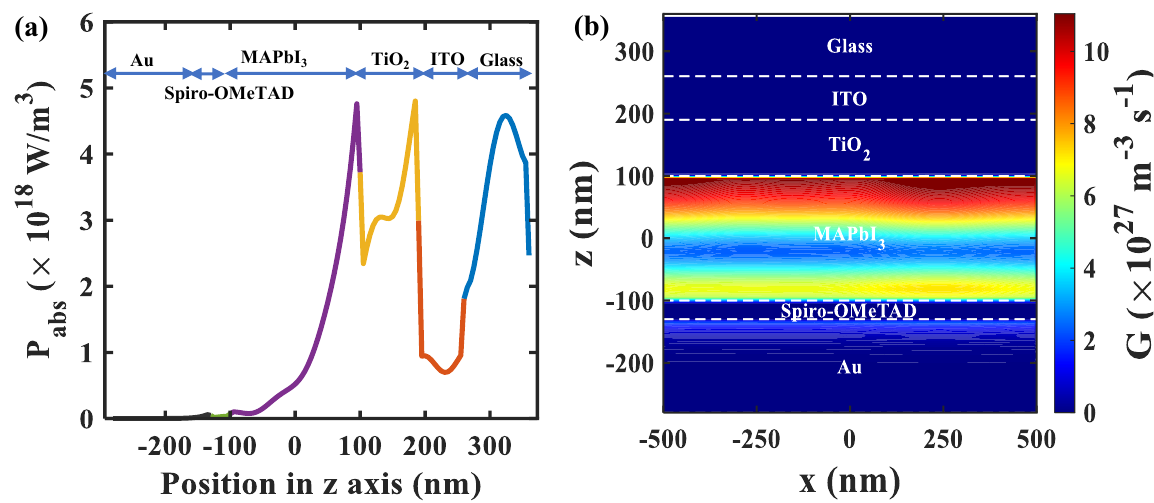}
\caption{(a) Total absorbed power density,
P\textsubscript{abs}, corresponding to the illumination of sunlight as a function of the vertical (z) position of the device. (b) Spatial profile of the photogeneration rate, G, in the x-z plane.}
\label{fig:generation}
\end{figure}

\autoref{fig:generation} shows how the optical redistribution caused by the hyperuniform front texture appears in the carrier-generation profile of the device. 
The absorbed power density, $\mathrm{P_{abs}}$, was mainly concentrated within the MAPbI$_3$ layer, while the adjacent transport and contact layers contributed much less to effective absorption as can be seen in \autoref{fig:generation}(a). 
The largest $\mathrm{P_{abs}}$ occurred at the TiO$_2$/MAPbI$_3$ interface and then decreased gradually toward the back of the absorber, which was consistent with the progressive attenuation of the incident photon flux as light propagates through the stack. 

The same behavior is reflected in the photogeneration profile in \autoref{fig:generation}(b). 
A broad, high-G region was observed across the MAPbI$_3$ layer, with the strongest generation near the front of the absorber and significant generation extending deeper into the active layer. 
This distribution follows Equations 
\ref{eq:Pabs}--\ref{eq:Gint} because the local generation rate was based on the absorbed optical power and thus reflected the local field intensity within the lossy perovskite. 
Compared to a purely planar response dominated by vertical thin-film interference, the hyperuniform nanoholes redirected part of the incident field into oblique and partially guided paths, which ultimately increased the optical path length inside MAPbI$_3$.
This enhancement originates from the prolonged confinement of the optical field within the glass and perovskite layer of the hyperuniform structure compared with the planar case, as further verified in Supplementary Video files.
Because the glass layer had a very low extinction coefficient at the operating wavelength, most of the trapped light in the glass eventually contributed to carrier generation in the perovskite region.
The field trapping effect was most effective near the red and near-band-edge regions, where single-pass absorption is weaker, and additional path-length enhancement directly contributes to photocurrent generation.

\subsection{Polarization-Independent Optical Absorption}

\begin{figure}[!htb]
\centering\includegraphics[width=\textwidth]
{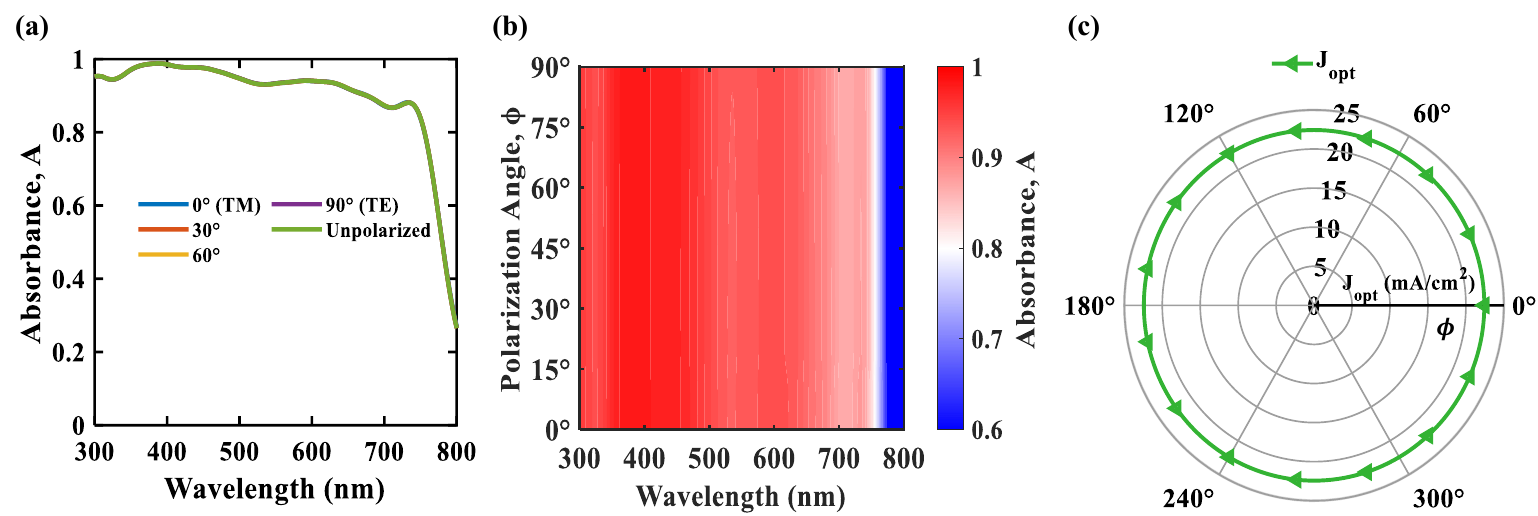}
\caption{Influence of light polarization on the optical absorption of the hyperuniform-patterned MAPbI$_3$ perovskite solar cell. (a) Simulated absorption spectra for both unpolarized light and various polarization angles ($\phi$), ranging from TM ($0^\circ$) to TE ($90^\circ$), showing nearly identical responses across the 300--800 nm range. (b) Absorption contour map as a function of wavelength and polarization angle, confirming a weak polarization dependence and a generally uniform optical response. (c) Polar plot of the optically calculated short-circuit current density, $\mathrm{J_{opt}}$, as a function of polarization angle, demonstrating an almost circular profile and thus a polarization-insensitive photocurrent generation.}
\label{fig:polarization}
\end{figure}

\autoref{fig:polarization} indicates that the optical response of the hyperuniform-patterned device changes very little as the polarization angle is rotated from TM ($0^\circ$) to TE ($90^\circ$). 
In \autoref{fig:polarization}(a), the absorbance spectra stayed nearly the same across most of the 300-800 nm wavelength range. 
The same behavior was seen more evidently in \autoref{fig:polarization}(b), where the contour map showed little variation along the polarization-angle axis. 
In other words, no distinct wavelength band becomes strongly polarization-selective within the simulated range. 
This weak sensitivity was also reflected in \autoref{fig:polarization}(c), where the polar plot of $\mathrm{J_{opt}}$ stayed close to circular, indicating that the integrated photocurrent changes only slightly with polarization angle.

We attribute this polarization-insensitive behavior to the absence of a preferred in-plane direction in the hyperuniform nanohole pattern.
In strongly anisotropic textures or one-dimensional gratings, only some selective polarization states can couple more efficiently than others because the available optical modes are direction-dependent. 
In our case, the statistically isotropic arrangement of the nanoholes spread scattering over many in-plane momentum channels; hence, both TM- and TE-like excitations experienced similar optical coupling conditions. 
As a result, the broadband absorbance and the corresponding $\mathrm{J_{opt}}$ remained nearly unchanged under polarization variation. 
This is desirable for practical solar-cell operation, since sunlight incident on the device is generally unpolarized.

\subsection{Incident Angle Tolerant Nanopattern}

\begin{figure}[!htb]
\centering\includegraphics[width=\textwidth]
{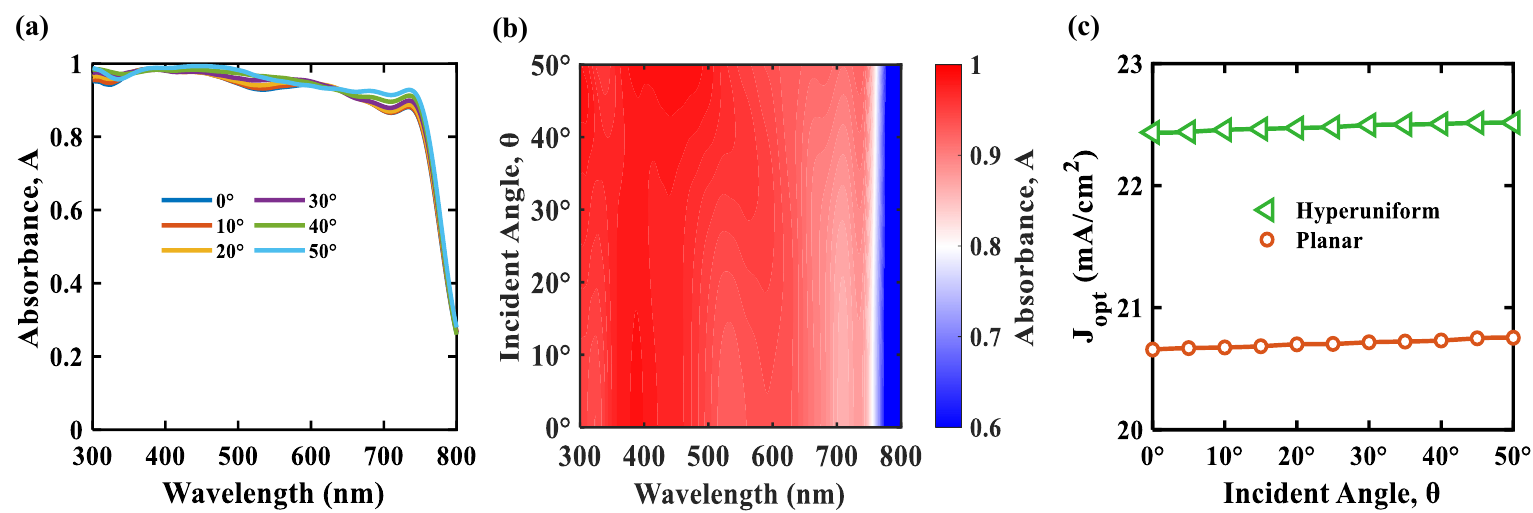}
\caption{Incident-angle dependence of broadband absorption and optical current generation in the hyperuniform-patterned MAPbI$_3$ perovskite solar cell. (a) Absorbance spectra calculated for illumination angles ($\theta$) between $0^\circ$ and $50^\circ$, indicating that the hyperuniform nanohole texture maintains a largely stable spectral response across most of the solar-relevant wavelength range, with noticeable deviations appearing mainly near the weakly absorbed long-wavelength edge at larger angles. (b) Wavelength--angle absorbance map of the hyperuniform structure, showing a smooth response without pronounced angle-selective resonances, consistent with robust broadband coupling under oblique incidence. (c) Comparison of $\mathrm{J_{opt}}$ versus incident angle for the hyperuniform and planar structures. The hyperuniform device maintains a high and nearly invariant $\mathrm{J_{opt}}$ across the entire angular range, while the planar reference remains lower. This demonstrates the superior angular stability of both photocurrent generation and enhancement enabled by the hyperuniform front texture.}
\label{fig:incident}
\end{figure}


To examine the angular robustness of the proposed texture, we evaluated the optical response of the hyperuniform-patterned device for incident angles from $0^\circ$ to $50^\circ$. This range was selected because simulations at larger angles became increasingly computationally demanding and showed convergence difficulties. The incident-angle analysis was performed using the BFAST source and boundary conditions, ensuring the injection angle remained well defined across the full simulated wavelength range.

As shown in \autoref{fig:incident}(a), the absorbance spectra remained largely stable over most of the 300--800 nm range as the illumination direction was tilted away from normal incidence. The differences between the spectra became more noticeable only at larger angles and were concentrated mainly near the long-wavelength edge, where MAPbI$_3$ already absorbed less efficiently in a single pass. Therefore, oblique illumination did not produce any severe spectral collapse in the main solar-flux region. The same conclusion was supported by the wavelength--angle contour map in \autoref{fig:incident}(b), which exhibited a smooth response without sharp angle-selective features. This behavior indicates that the optical enhancement was not governed by a narrowly tuned angular resonance, but instead was preserved over a broad range of illumination directions.

\autoref{fig:incident}(c) further shows that the optically calculated short-circuit current density, $\mathrm{J_{opt}}$, of the hyperuniform device remained nearly unchanged from $0^\circ$ to $50^\circ$. More importantly, the hyperuniform structure maintained a consistently higher $\mathrm{J_{opt}}$ than the planar reference across the full angular range considered here. Thus, the advantage of the textured front glass was not limited to normal incidence, but persisted under oblique illumination as well. This result suggests that the hyperuniform nanohole layer preserved efficient optical coupling even when the incident wave vector acquired a larger in-plane component.

The physical origin of this behavior is consistent with the hyperuniform pattern's statistical isotropy. As discussed previously for the polarization study, the absence of a preferred in-plane direction allowed the nanohole layer to redistribute the incident field into a broad set of in-plane momentum channels rather than coupling through only a narrow set of angle-sensitive pathways. Consequently, the structure sustained coupling into oblique and partially guided optical modes over a wide angular window. In addition, oblique rays traversed a longer geometric path through the multilayer stack than normally incident rays. If optical coupling remained effective, this longer path could still support strong absorption, especially in the weakly absorbed long-wavelength region.

By contrast, the planar reference remained optically inferior throughout the same angular range because it relied primarily on specular thin-film interference and did not provide comparable lateral momentum redistribution. As a result, although the planar device also showed a relatively smooth angular trend, it could not access the same enhanced light-trapping channels as the hyperuniform structure. Overall, \autoref{fig:incident}(a)--(c) demonstrates that hyperuniform front-glass patterning improved not only broadband absorption, but also the angular stability of photocurrent generation, which is highly desirable for realistic outdoor photovoltaic operation.

\subsection{Absorption Tuning through Nanohole Radius Variation}


\begin{figure}[!htb]
\centering\includegraphics[width=\textwidth]
{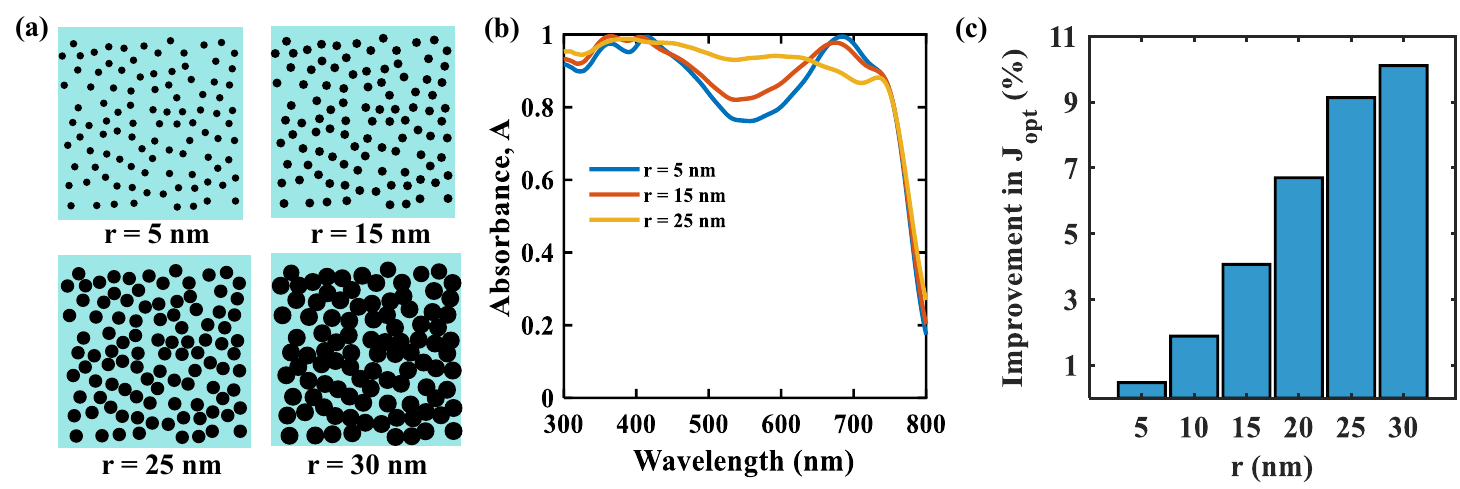}
\caption{Dependence of the hyperuniform nanohole morphology and optical response on nanohole radius. (a) Top-view renderings of the hyperuniform texture for r = 5, 15, 25, and 30 nm, showing the progressive reduction in inter-hole spacing as the radius increases while the nanohole-center coordinates are kept fixed. At r = 30 nm, neighboring features begin to overlap, indicating the onset of morphological degradation of the intended isolated-hole topology. (b) Simulated absorbance spectra for r = 5, 15, and 25 nm. Increasing the radius strengthens broadband absorption and suppresses interference-related spectral undulations, particularly toward the longer-wavelength region. (c) Relative enhancement in the optically calculated short-circuit current density, $\mathrm{J_{opt}}$, with respect to the planar reference as a function of nanohole radius. The enhancement increases nearly monotonically with radius, while r = 25 nm represents the largest non-overlapping geometry considered practical in the present design.}
\label{fig:radius}
\end{figure}

\autoref{fig:radius} shows how the nanohole radius sets both the geometric features of the hyperuniform texture and the strength of the resulting optical perturbation. 
In \autoref{fig:radius}(a), increasing the radius, r, while keeping the nanohole centers fixed progressively narrowed the spacing between adjacent holes. 
For small radii, the patterned glass was only weakly perturbed relative to the planar interface. 
At r = 25 nm, the texture became substantially stronger, while the holes remained distinct. 
When the radius was increased further to r = 30 nm, neighboring holes began to merge, indicating that the morphology was no longer consistent with the intended isolated-hole hyperuniform pattern. 
Thus, the practically relevant design space was limited to r $\leq$ 25 nm.

The spectral response for different nanohole radii was illustrated in \autoref{fig:radius}(b). 
For r = 5 nm, the absorbance stayed close to that of a weakly textured surface and retained more visible oscillatory structure. 
Increasing the radius to 15 nm, then to 25 nm, raised the absorbance across a broad portion of the 300-800 nm wavelength range, smoothing the spectral variation. 
The effect was most effective toward the red and near-band-edge region, where absorption in the thin MAPbI$_3$ layer is less efficient in a single pass. 
A larger radius introduced a stronger refractive-index perturbation at the front surface, which broadened the scattering response and couples more of the incident field into oblique propagation paths within the stack. 
As a result, the optical path length inside the absorber increased, and the response became less dominated by narrow interference features.

\autoref{fig:radius}(c) confirms that these spectral changes translate into a systematic gain in the integrated optical current. 
The relative enhancement in $\mathrm{J_{opt}}$ increased nearly monotonically with radius, consistent with progressively stronger light trapping as the surface texture becomes more pronounced. 
Although the largest calculated gain occurred at r = 30 nm, the overlap visible in \autoref{fig:radius}(a) indicated that this case no longer preserved the intended nanohole topology. 
Therefore, r = 25 nm was selected as the optimal practical radius because it provides strong broadband absorption enhancement and a substantial improvement in $\mathrm{J_{opt}}$ without compromising the isolated nanohole array's geometric integrity.

\subsection{Role of Nanohole Height in Absorption Enhancement}


\begin{figure}[!htb]
\centering\includegraphics[width=\textwidth]
{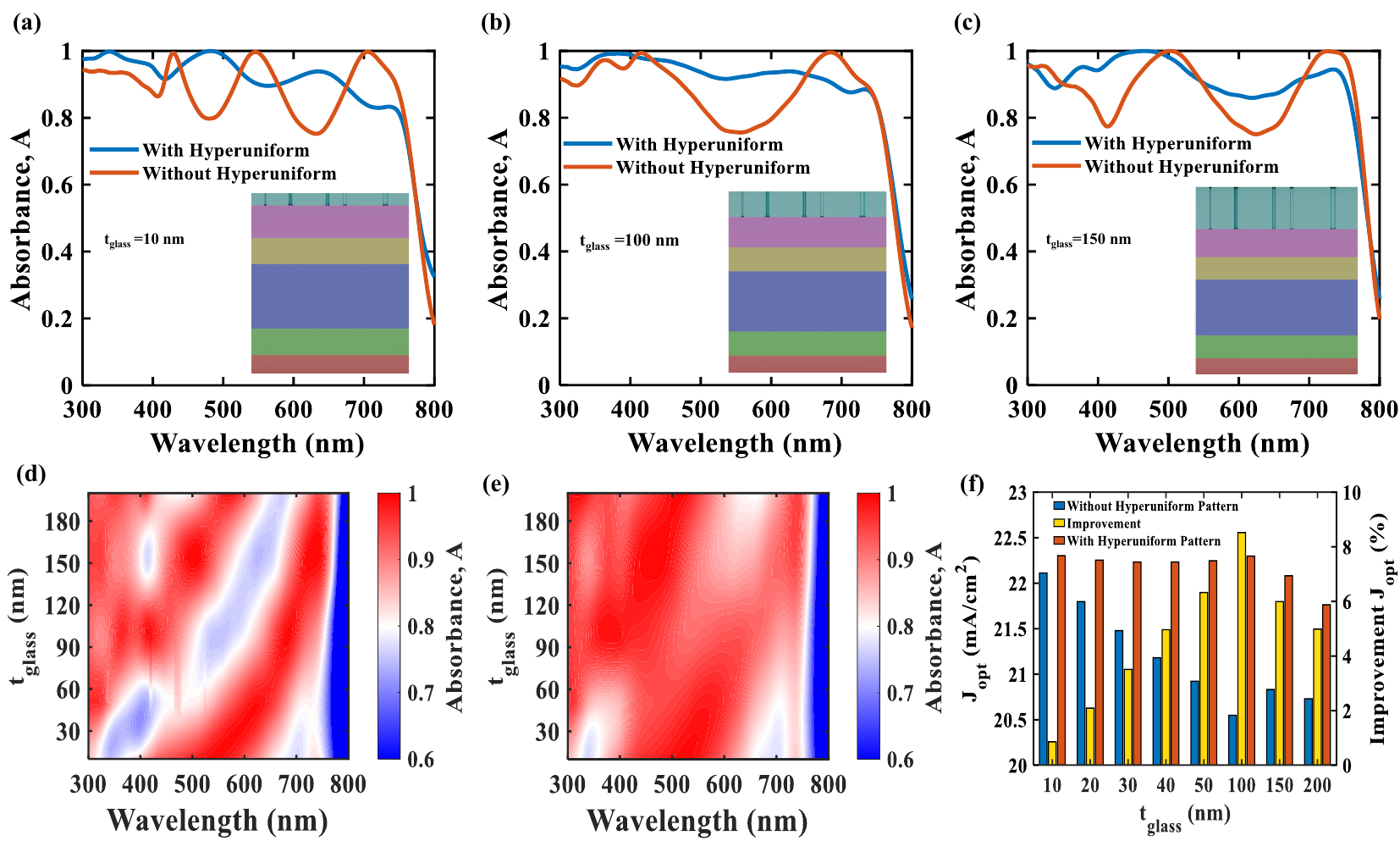}
\caption{Influence of nanohole height, $\mathrm{t_{glass}}$, on the optical absorption of the perovskite solar cell for a fixed nanohole radius of r = 25 nm and hyperuniform nanohole center positions. (a--c) Absorption spectra for $\mathrm{t_{glass}} = 10$, 100, and 150 nm, comparing the hyperuniform-patterned device with the unpatterned reference. (d,e) Absorption contour maps as a function of wavelength and patterned-layer thickness for the unpatterned and hyperuniform cases, respectively. The hyperuniform layer suppressed sharp interference features and resonant fluctuations, leading to a smoother broadband absorption response across the visible and near-infrared regions. (f) Optically calculated short-circuit current density, $\mathrm{J_{opt}}$, for the unpatterned and hyperuniform devices, together with the relative improvement in $\mathrm{J_{opt}}$ as a function of nanohole height. The largest enhancement is obtained at $\mathrm{t_{glass}} = 100$ nm.}
\label{fig:height}
\end{figure}

Considering previous nanohole radius findings, we examined the influence of the nanohole pattern height, $\mathrm{t_{glass}}$, on the solar cell's optical response.
The absorbance spectra in \autoref{fig:height}(a--c) illustrate that changing $\mathrm{t_{glass}}$ alters the absorption of both structures, but the response of the unpatterned device is much more oscillatory. 
In the planar structure, pronounced maxima and minima were observed across the visible and near-band-edge regions, and these spectral features shifted as the glass thickness was varied. 
This behavior is characteristic of thickness-dependent Fabry--P\'erot interference, where the optical phase accumulated in the front dielectric determines the standing-wave distribution inside the multilayer stack. 
As that phase condition changed, the field antinodes moved relative to the MAPbI$_3$ layer, so the overlap between the optical field and the absorber became strongly wavelength- and thickness-dependent.

However, the hyperuniform-patterned device behaved differently from planar counterparts of similar thickness. 
For $\mathrm{t_{glass}} =$ 10, 100, and 150 nm, the absorption remained comparatively smooth, and the sharp oscillatory structure seen in the unpatterned reference was substantially reduced, as shown in \autoref{fig:height}. 
The difference was especially relevant in the longer-wavelength region, where single-pass absorption in the perovskite was weaker, and the device benefited more from additional light trapping. 
This behavior was observed because the hyperuniform nanoholes do not function as a single resonant texture.
Instead, they scattered the incident wave into a broad distribution of in-plane momentum states, which promoted coupling into oblique and partially guided optical paths. 
Since the optical power was redistributed over many channels rather than being tied mainly to the particular cavity mode, the overall absorption became less sensitive to the exact front-glass thickness.

This interpretation was reinforced by the wavelength-thickness maps in \autoref{fig:height}(d,e). 
The unpatterned structure in \autoref{fig:height}(d) exhibited distinct diagonal interference bands, indicating that the absorption varied strongly as the optical thickness of the front dielectric changed. 
By contrast, the hyperuniform case in \autoref{fig:height}(e) showed a much more uniform absorption landscape, with visibly weaker resonance stripes over the same thickness range. 
Typically, the patterned interface breaks the lateral symmetry of the planar entrance surface and relaxes the strict momentum constraint of normal-incidence coupling. 
As a result, the incoming field could access a broader region of optical phase space, reducing the dominance of narrow, thickness-specific resonances and increasing the effective path length within the absorber.

The integrated consequence of this behavior is summarized in \autoref{fig:height}(f). 
The optically calculated short-circuit current density, $\mathrm{J_{opt}}$, was higher for the hyperuniform structure over the full range of $\mathrm{t_{glass}}$, and the relative improvement remains positive throughout the sweep. 
The largest enhancement occurred at $\mathrm{t_{glass}} =$ 100 nm, suggesting that an intermediate patterned-layer height provided the most favorable balance between scattering strength and optical coupling. 
When the patterned layer was too shallow, the refractive-index perturbation was comparatively weak and only a limited fraction of the incident field was redirected. 
When the layer became thicker, the texture still improved absorption. Still, part of the added optical interaction occurred within the front glass itself, and the incremental gain in absorber coupling no longer increases. 
Overall, these results demonstrated that hyperuniform patterning contributed not only to enhanced broadband absorption but also to reduced sensitivity of the optical response to front-glass thickness, which is advantageous for a fabrication process where this dimension may vary between samples.

\subsection{Benchmarking against Conventional Uniform Nanohole Arrays}

\begin{figure}[!htb]
\centering\includegraphics[width=\textwidth]
{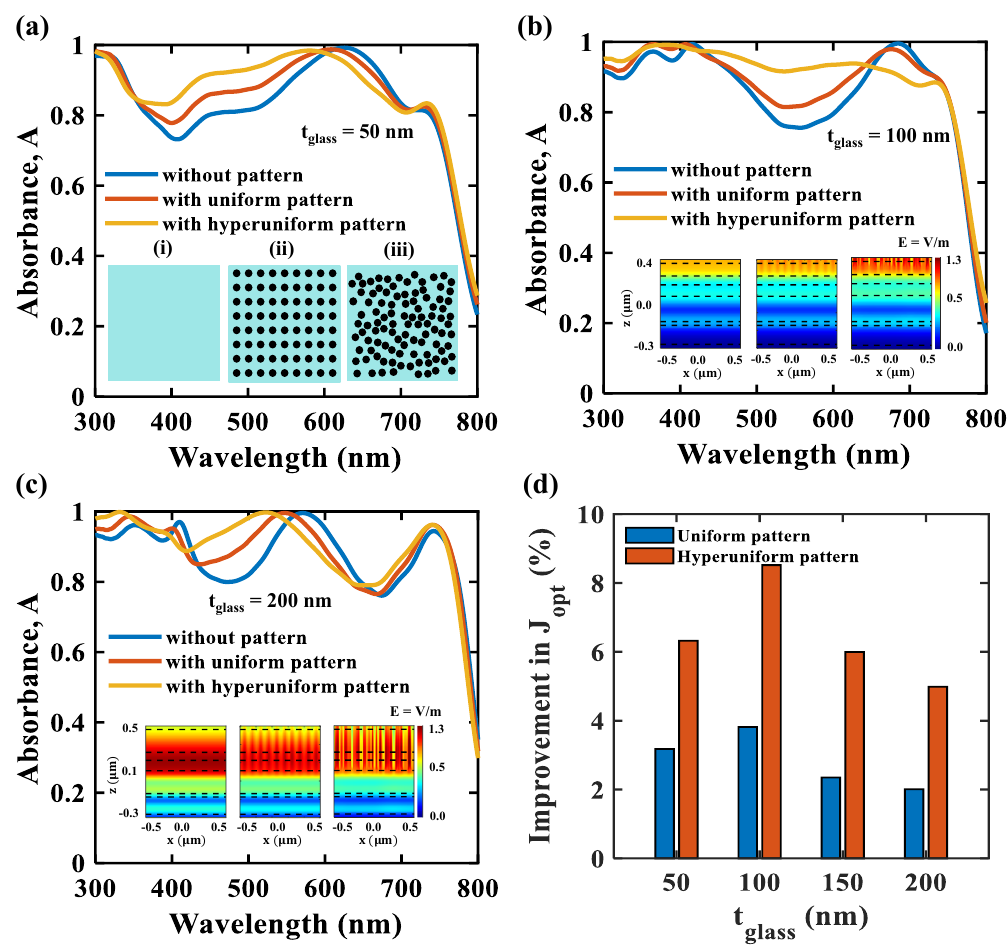}
\caption{Comparison of optical absorption for perovskite solar cells incorporating the nanohole-textured top glass layer with three surface configurations: unpatterned glass, a periodic (uniform) nanohole array, and a hyperuniform nanohole array (inset). (a--c) Absorption spectra for nanohole heights of 50, 100, and 200~nm, respectively, highlighting the broadband absorption enhancement obtained with the hyperuniform pattern relative to both the unpatterned and periodic cases. The insets in (b-c) show the electric field profile at a wavelength of 540 nm. (d) Optically calculated short-circuit current density, $\mathrm{J_{sc}}$ as a function of nanohole layer thickness for the three configurations, confirming that the hyperuniform nanohole layer provides the largest $\mathrm{J_{sc}}$ across the investigated thickness range.}
\label{fig:uniform_vs_hyperuniform}
\end{figure}

To isolate the role of point-pattern statistics, we compared three top-glass configurations: an unpatterned planar interface, a periodic nanohole array, and a hyperuniform nanohole array.
\autoref{fig:uniform_vs_hyperuniform}(a)--(c) shows that all three structures behave similarly at the short-wavelength end of the spectrum, where MAPbI$_3$ already absorbed strongly in a single pass. 
The differences become much more visible at longer wavelengths, especially in the red and near-band-edge region, where light trapping is more important. 
For each investigated patterned-layer height, the hyperuniform texture maintained the highest absorbance over a broad part of the 300--800 nm window. 
The periodic array also improved absorption relative to the planar reference, but its spectra exhibited more pronounced peaks and valleys, indicating greater sensitivity to wavelength-specific optical coupling conditions.

In the periodic case, the texture introduced a discrete set of in-plane momentum components, so coupling into lateral or partially guided optical states was governed by a limited number of reciprocal-lattice channels. 
As a result, the absorption enhancement was tied more strongly to specific phase-matching conditions, which produced the oscillatory line shape seen in \autoref{fig:uniform_vs_hyperuniform}(a)--(c). 
By comparison, the hyperuniform arrangement did not impose long-range translational order, yet it still preserved short-range spatial correlation. 
That combination spread the scattering response over a broader range of in-plane wave vectors and reduced the dominance of narrow diffractive resonances. 
The result was a smoother, broader redistribution of the optical field into the absorber. In practical terms, the hyperuniform layer was better able to increase the effective optical path length in MAPbI$_3$ without concentrating the gain into only a few isolated spectral bands.

\autoref{fig:uniform_vs_hyperuniform}(d) shows that these spectral differences persist after solar weighting and integration into the $\mathrm{J_{opt}}$. 
The hyperuniform design yielded the largest improvement at every nanohole height considered, while the periodic pattern laid between the hyperuniform and planar cases. 
The advantage was most evident once the patterned region became sufficiently deep to provide strong index modulation, but it did not increase without limit. 
At very small heights, the perturbation was too weak to redirect a large fraction of the incident field. At higher heights, scattering remained strong, but not all of the additional optical interactions were efficiently converted into useful absorption within the perovskite. 
An intermediate patterned-layer height, therefore, offered the most favorable balance between scattering strength and absorber coupling. 
Overall, this benchmark indicates that the benefit of the hyperuniform design is not simply due to the presence of nanoholes; it arises from the texture's specific spatial statistics, which enable broadband light trapping more effectively than a conventional periodic array.

\subsection{Robustness of Device Functionality to Nanohole Fabrication Imperfections}

\begin{figure}[!htb]
\centering\includegraphics[width=\textwidth]
{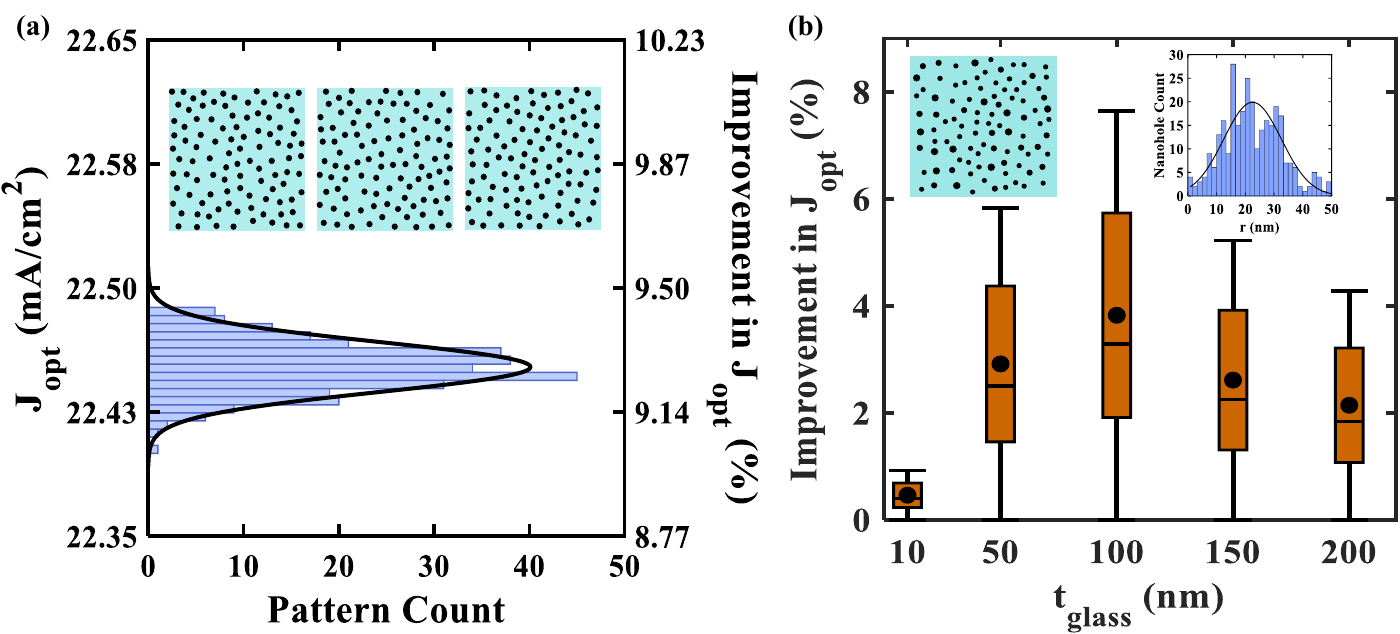}
\caption{Fabrication-tolerance analysis of the hyperuniform nanohole glass layer under geometric disorder. (a) Histogram of the optically calculated short-circuit current density, $\mathrm{J_{opt}}$, obtained from multiple independently generated hyperuniform pattern realizations. The narrow distribution indicates that realization-to-realization fluctuations in the nanohole arrangement introduced only a limited variation in the optical current. The right axis shows the corresponding relative improvement in $\mathrm{J_{opt}}$ with respect to the planar reference, confirming that the optical enhancement remained tightly distributed around a positive mean value. (b) Box-and-whisker distributions of the relative improvement in $\mathrm{J_{opt}}$ for $\mathrm{t_{glass}}=$~10, 50, 100, 150, and 200~nm when both the hyperuniform pattern and nanohole radius are allowed to vary stochastically. The black dots denote the mean values. The largest median enhancement was observed near $\mathrm{t_{glass}}=$~100~nm, although this case also exhibits the widest spread, indicating greater sensitivity of the optimum configuration to fabrication-induced perturbations. In contrast, the $10$~nm structure shows the narrowest distribution but only a marginal optical gain. The persistence of predominantly positive distributions across all investigated thicknesses demonstrates that the hyperuniform texture retains its optical advantage under fabrication-relevant disorder. Insets: representative hyperuniform nanohole realization (left) and distribution of nanohole radii used to emulate dimensional fluctuations (right), with the solid line representing the corresponding Gaussian fit.}
\label{fig:fab}
\end{figure}


To evaluate the fabrication tolerance of the hyperuniform nanohole layer, we introduced geometric disorder at two levels. 
First, multiple hyperuniform point patterns were generated using the same minimum inter-hole spacing criterion in order to capture realization-to-realization variations in the nanohole arrangement. 
Second, stochastic variations in the nanohole radius were applied to emulate fabrication-induced dimensional errors. 
The resulting radius values, shown in the inset of \autoref{fig:fab}(b), follow an approximately Gaussian distribution centered around the nominal design radius (25 nm).
To maintain a physically meaningful geometry, radius realizations that produced overlapping neighboring holes were excluded. 
As a result, the final radius ensemble deviated slightly from an ideal Gaussian to represent a practically manufacturable structure.

The effect of pattern realization alone is first summarized in \autoref{fig:fab}(a). The histogram of $\mathrm{J_{opt}}$ remained narrowly distributed, indicating that fluctuations in the hyperuniform point pattern produced only a limited variation in the solar-weighted optical current. 
The secondary axis further shows that this narrow spread corresponded to a consistent improvement relative to the planar reference. 
This indicates that the optical benefit did not depend on one exceptional realization of the hyperuniform layout; instead, it was preserved across multiple statistically equivalent patterns.

The combined influence of pattern variation and nanohole-radius disorder is presented in \autoref{fig:fab}(b) for $\mathrm{t_{glass}}=$ 10, 50, 100, 150, and 200 nm. 
For all investigated thicknesses, the box-and-whisker distributions remained predominantly above zero, which confirms that the hyperuniform nanohole layer retained its optical advantage even after fabrication-relevant perturbations were introduced.
The black dots indicate the mean values, while the horizontal lines inside the boxes represent the medians. 
Among all cases, the highest central tendency was obtained for $\mathrm{t_{glass}}=$~100 nm, in agreement with the optimum thickness identified previously for the ideal structure. 
However, this same case also exhibited the widest spread, implying that the geometry delivering the strongest optical gain was also somewhat more sensitive to fabrication-induced perturbations. 
By contrast, the $\mathrm{t_{glass}}=$~10 nm case showed the narrowest distribution but only a small improvement, indicating that a very shallow patterned layer did not provide sufficient refractive-index modulation and scattering strength to maximize light trapping. 
The intermediate and larger thicknesses, especially $\mathrm{t_{glass}}=$~50, 150, and 200 nm, remained clearly beneficial, although their gains were lower than that of the $\mathrm{t_{glass}}=$~100 nm case.

This behavior is physically consistent with the broadband operating principle of the hyperuniform texture. 
If the enhancement mainly stemmed from a narrowly tuned diffractive resonance, relatively minor geometric perturbations would have caused significant detuning and a substantial collapse in $\mathrm{J_{opt}}$. 
Instead, the positive distributions observed across all thicknesses indicate that the hyperuniform nanohole layer redistributed optical momentum through a broad set of in-plane scattering channels. Hence, local variations in hole radius altered the strength of the response without destroying the underlying light-trapping mechanism. 
In this sense, fabrication disorder affected the magnitude of the enhancement more strongly than its retention.

Overall, \autoref{fig:fab} showed that the proposed hyperuniform nanohole architecture was reasonably tolerant to fabrication-induced variations in both pattern realization and nanohole radius. Although the optimum configuration near $\mathrm{t_{glass}}=100$ nm was somewhat more statistically dispersed than the shallower cases, the enhancement remained positive throughout the investigated thickness range. From a practical device-fabrication perspective, this result suggests that the predicted optical benefit did not require unrealistically precise control over every individual nanohole, provided that the global hyperuniform character of the texture was preserved.

\subsection{Enhancement in Electrical Performance via Hyperuniform Patterning}

\begin{table}[ht]
\centering
\setlength{\tabcolsep}{10pt} 
\renewcommand{\arraystretch}{1.2} 

\begin{tabular}{lcccc}
\hline
Pattern & $\mathbf{V_{oc}}$ \textbf{(V)} & $\mathbf{J_{sc}}$ $\mathbf{(mAcm^{-2})}$ & $\mathbf{Fill}$ $\mathbf{ Factor}$ \textbf{(\%)} & $\mathbf{PCE}$ \textbf{(\%)} \\
\hline
No Pattern & 1.12 & 21.57 & 87.00 & 21.03 \\

Uniform & 1.13 & 23.37 & 82.05 & 21.65 \\

Hyperuniform & 1.13 & 23.92 & 87.66 & 23.62 \\
\hline
\end{tabular}

\caption{Electrical Performance Enhancement of Different Patterning Schemes}
\label{tab:elec_res}
\end{table}

\begin{figure}[!htb]
\centering\includegraphics[width=\textwidth]{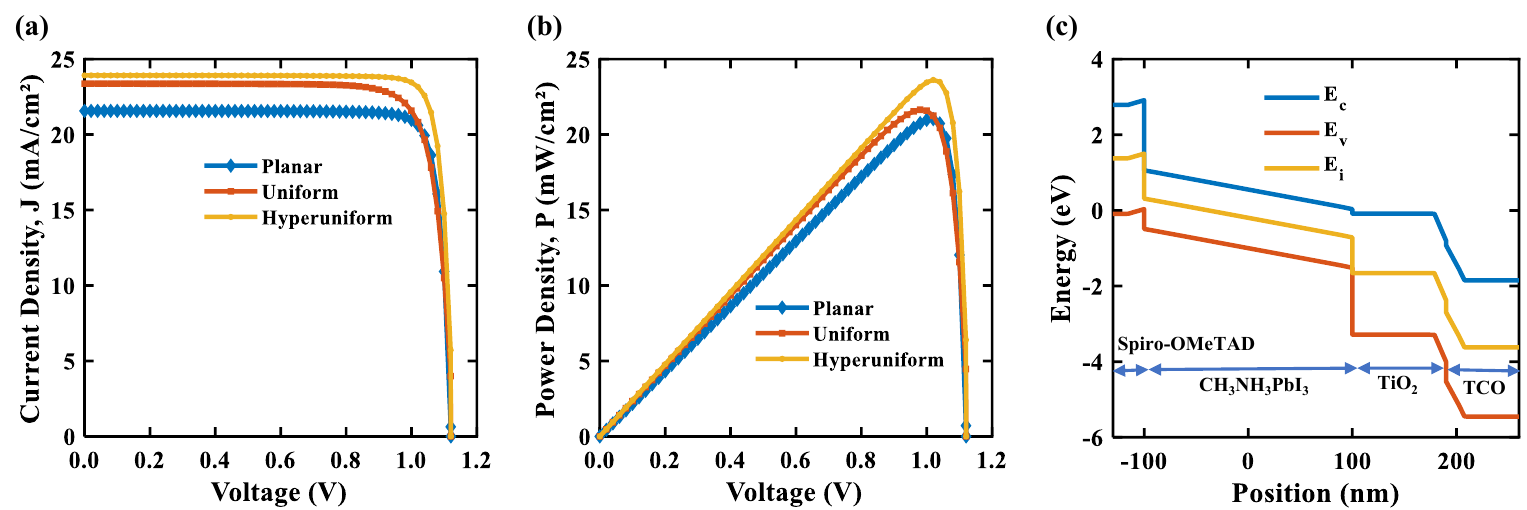}
\caption{Electrical performance comparison of planar, periodic, and hyperuniform MAPbI$_3$ perovskite solar cells obtained from drift--diffusion simulations using the optically calculated generation profiles. (a) Simulated illuminated current density--voltage (J--V) characteristics, showing the higher photocurrent delivered by the hyperuniform device across the operating-voltage range. (b) Corresponding power density--voltage (P--V) characteristics, where the hyperuniform structure exhibits the largest maximum power output, consistent with its superior power-conversion efficiency. (c) Energy-band diagram of the Glass/ITO/TiO$_2$/MAPbI$_3$/Spiro-OMeTAD/Au device architecture used in the electrical simulation, indicating the band alignment responsible for electron extraction through TiO$_2$ and hole extraction through Spiro-OMeTAD.}
\label{fig:JV}
\end{figure}

\autoref{tab:elec_res} and \autoref{fig:JV} show that the optical advantages of front-glass nanopatterning were retained after the optically calculated generation profiles were coupled to drift--diffusion transport. Among the three investigated configurations, the hyperuniform device delivered the best overall electrical performance, with $\mathrm{V_{oc}} =$~1.13~V, $\mathrm{J_{sc}} =$~23.92~mA/cm$^2$, $\mathrm{FF} =$ 87.66\%, and $\eta =$ 23.62\%. By comparison, the planar reference yielded $\mathrm{V_{oc}} =$~1.12~V, $\mathrm{J_{sc}} =$~21.57~mA/cm$^2$, $\mathrm{FF} =$ 87.00\%, and $\eta =$ 21.03\%, whereas the uniform nanohole structure reached $\mathrm{V_{oc}} =$~1.13~V, $\mathrm{J_{sc}} =$~23.37~mA/cm$^2$, $\mathrm{FF} =$~82.05\%, and $\eta =$~21.65\%. These results indicate that the dominant photovoltaic improvement originated from the increase in $\mathrm{J_{sc}}$, while the change in $\mathrm{V_{oc}}$ remained comparatively small.

This trend is consistent with the nanostructure's position in the solar stack. Since the nanohole pattern was introduced into the front glass rather than into the electronically active junction, it primarily modified optical coupling and carrier generation, not the device's band alignment. As a result, both patterned structures generated more photocarriers in the MAPbI$_3$ layer, especially in the weakly absorbed long-wavelength region, which increased $\mathrm{J_{sc}}$. By contrast, $\mathrm{V_{oc}}$ changed only slightly because it depends mainly on the balance between generation and recombination. In simplified form, $\mathrm{V_{oc}} \approx \frac{\mathrm{kT}}{\mathrm{q}}\ln\!\left(\frac{\mathrm{J_{ph}}}{\mathrm{J_0}}+1\right)$, hence, a moderate increase in photocurrent generally produces only a limited voltage gain when $\mathrm{J_0}$ remains nearly unchanged.

However, the fill factor appeared as a performance-limiting factor in the case of uniform distribution. Although the uniform nanohole structure improved $\mathrm{J_{sc}}$ relative to the planar device, its $\mathrm{FF}$ decreased substantially, which limited the final gain in $\eta$. This reduction was most likely not caused by a change in the junction energetics, since \autoref{fig:JV}(c) shows that the TiO$_2$/MAPbI$_3$/Spiro-OMeTAD extraction pathway remained unchanged. Instead, the lower $\mathrm{FF}$ of the uniform case likely originated from the nature of its optical enhancement. As discussed earlier in \autoref{fig:uniform_vs_hyperuniform}, the periodic texture redistributed light through a limited number of discrete reciprocal-lattice channels and therefore produced a more resonance-selective optical response. Such a response can generate a less spatially and spectrally uniform photocarrier distribution, which in turn may strengthen local carrier-density gradients and bias-dependent recombination losses near the maximum-power operating region. In practical terms, the periodic structure increased photocurrent, but it did not preserve the squareness of the $\mathrm{J}$--$\mathrm{V}$ curve as effectively as the hyperuniform texture.

By contrast, the hyperuniform structure maintained both the highest $\mathrm{J_{sc}}$ and a high $\mathrm{FF}$, which indicates a more favorable balance between light trapping and carrier extraction. Since the hyperuniform arrangement redistributed the incident field across a broader, more isotropic set of in-plane momentum channels, its optical enhancement was less concentrated in narrowly tuned resonant pathways. The resulting generation profile was therefore expected to remain more broadband and electrically benign, allowing the device to benefit from stronger long-wavelength absorption without incurring the pronounced $\mathrm{FF}$ penalty observed in the uniform case. This interpretation is consistent with \autoref{fig:JV}(a), where the hyperuniform device sustained the highest current density across most of the operating-voltage range, and with \autoref{fig:JV}(b), where it exhibited the largest maximum power output.

Overall, the results demonstrate that the hyperuniform nanohole layer improved the device more effectively than either the planar or uniform counterparts. The front-glass texture increased carrier generation while preserving the essential band structure of the planar heterojunction, and the hyperuniform statistics enabled that optical gain to translate more efficiently into usable electrical power. Consequently, the hyperuniform design achieved the highest simulated photovoltaic performance among the three cases considered here.

\subsection{Proposed Fabrication Process}

\begin{figure}[!htb]
\centering\includegraphics[width=5.8in]
{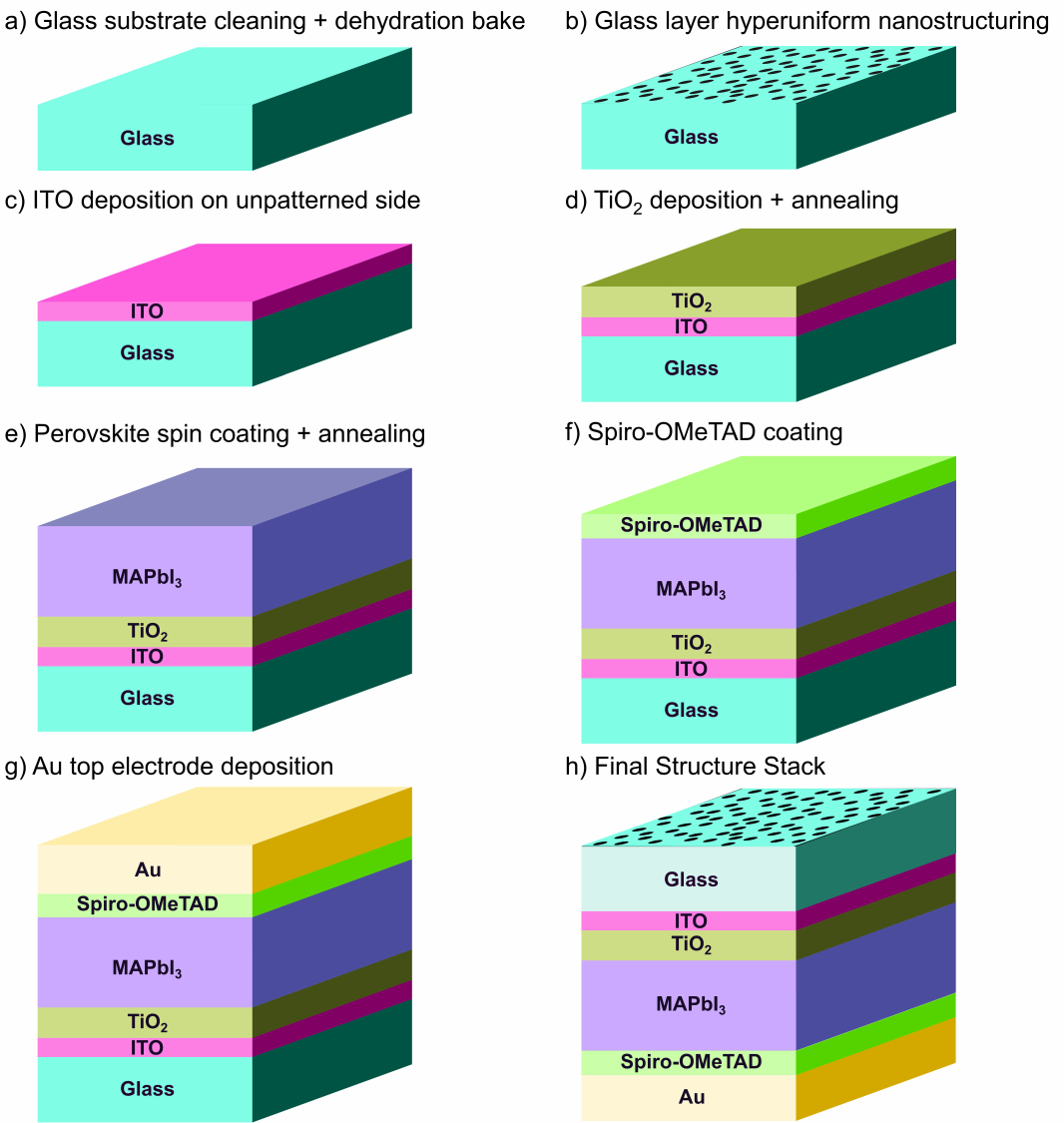}
\caption{Proposed fabrication route for the hyperuniform-glass perovskite solar-cell architecture considered in this work. The process comprises (a) cleaning and dehydration treatment of the bare glass substrate; (b) hyperuniform nanostructuring of the glass through hard-mask deposition, e-beam resist coating, electron-beam lithography, resist development, pattern transfer, glass etching, and mask removal; (c) deposition of the ITO transparent electrode on the unpatterned side of the substrate; (d) TiO$_2$ deposition and annealing; (e) MAPbI$_3$ spin coating and annealing; (f) Spiro-OMeTAD deposition; (g) Au top-electrode deposition; and (h) the final Glass/ITO/TiO$_2$/MAPbI$_3$/Spiro-OMeTAD/Au device on a hyperuniform-patterned glass substrate.}
\label{fig:fab_process}
\end{figure}

\autoref{fig:fab_process} presents a proposed fabrication sequence for realizing the hyperuniform nanohole-patterned PSC architecture. 
The process provides a physically plausible route for experimental implementation of the simulated structure.
Fabrication starts on a glass substrate, which is first cleaned and dehydrated to remove organic and particulate contaminants and improve the uniformity of subsequent coating steps. 
A thin hard mask is then deposited on the glass surface, followed by coating of an electron-beam resist. The hyperuniform nanohole pattern is subsequently defined by electron-beam lithography, which is well suited to the sub-50 nm feature sizes considered here. 
After exposure, the resist is developed to form the desired nanohole array, and the pattern is transferred into the underlying hard mask by an intermediate etching step. 
The patterned hard mask is then used to etch the glass surface, thereby producing the hyperuniform nanostructured texture. 
Finally, the residual resist and hard-mask layers are removed to obtain a clean patterned glass substrate \cite{Bu2004SciRep, Chen2019JAC}. This process is detailed in Figure S4 of the Supplementary Information.

Following the formation of the hyperuniform texture, the transparent conductive oxide is introduced. 
Here, ITO is deposited on the unpatterned side of the glass substrate, leaving the textured glass surface exposed on the incident side, while the active photovoltaic stack is formed on the opposite face. 
This arrangement is consistent with the optical role of the hyperuniform pattern as a light-management feature and avoids unnecessary complexity associated with depositing the full device stack directly onto a nanostructured surface. 
After ITO deposition, a compact TiO$_2$ layer is formed on the ITO-coated side to serve as the electron-transport layer. 
This TiO$_2$ film may be deposited by a standard thin-film route such as sputtering or solution processing, followed by annealing to improve film densification, interfacial quality, and electron-extraction characteristics \cite{Chen2015SciRep}.

The perovskite absorber is then deposited on top of the TiO$_2$ layer. 
In the current configuration, MAPbI$_3$ is treated as the photoactive material and is assumed to be formed via a conventional solution-based deposition process, such as spin coating, followed by thermal annealing. 
This postdeposition annealing step is important for promoting solvent removal, crystallization, and formation of a continuous absorber layer with suitable optoelectronic quality \cite{Jeon2014NatMat}.
After the absorber layer is formed, Spiro-OMeTAD is deposited as the hole-transport layer. 
This layer facilitates hole extraction from the perovskite and provides an appropriate interface for depositing the metallic back contact. 
The device is then completed by deposition of a Au top electrode, typically by thermal evaporation, leading to the final Glass/ITO/TiO$_2$/MAPbI$_3$/Spiro-OMeTAD/Au architecture \cite{Chen2015SciRep}.


\subsection{Comparative Analysis}

\autoref{tab:nanotexture_comparison} compares the current hyperuniform front-glass design with several recent nanotextured PSC studies that achieved similar efficiency levels, but through significantly different structural approaches. These reports can be interpreted more clearly by dividing them into two categories: externally added optical textures, which mainly function through photon management, and textures incorporated directly into electronically active regions, where optical improvement is linked to changes in crystallization, defect density, or charge-transport behavior.

Among the most relevant external light-management approaches, Tockhorn \textit{et al.} \cite{Tockhorn2025ACP} combined shallow front nanotextures with a NaF antireflection coating and reported $\mathrm{J_{sc}}=23.10$~mAcm$^{-2}$, and $\mathrm{PCE=19.70}\%$, corresponding to a relative PCE improvement of $5.3\%$. Similarly, Wang \textit{et al.} \cite{Wang2021OptLett} used a holographic-lithography-derived moth-eye antireflection film and obtained $\mathrm{J_{sc}=23.20}$~mAcm$^{-2}$ and $\mathrm{PCE=20.27}\%$, with a PCE gain of $4.8\%$. 
In both cases, the optical benefit originated mainly from reflection suppression and front-side light coupling. 
The current hyperuniform design is most similar to these studies because the nanostructure is also kept away from the electronically active junction. 
However, the current device achieved a significantly higher overall efficiency of $23.62\%$ with $\mathrm{J_{sc}=23.92}$~mAcm$^{-2}$, and a much higher FF of $87.66\%$, showing an increase of 12.3\% relative improvement.
This suggests that front-substrate hyperuniform patterning can deliver a stronger optical benefit than conventional periodic antireflection texturing while remaining electrically benign.

\begin{table*}[t]
\caption{Comparison of recent nanotextured light-management strategies in perovskite solar cells with power conversion efficiencies comparable to the present hyperuniform nanohole study.}
\label{tab:nanotexture_comparison}
\centering
\scriptsize
\setlength{\tabcolsep}{3pt}
\renewcommand{\arraystretch}{1.15}

\begin{tabularx}{\textwidth}{
Y
Y
C{0.90cm}
C{1.25cm}
C{0.90cm}
C{0.90cm}
C{2cm}
C{0.8cm}
}
\toprule
\textbf{Nanotexture} &
\textbf{Solar Cell Stack} &
\makecell[c]{\textbf{$\mathrm{V_{oc}}$} \\ \textbf{(V)}} &
\makecell[c]{\textbf{$\mathrm{J_{sc}}$} \\ \textbf{(mAcm$^{-2}$)}} &
\makecell[c]{\textbf{FF} \\ \textbf{(\%)}} &
\makecell[c]{\textbf{PCE} \\ \textbf{(\%)}} &
\makecell[c]{\textbf{Relative PCE} \\ \textbf{Gain (\%)}}  &
\textbf{Ref.} \\
\midrule

Hexagonal sinusoidal front texture &
Glass\slash NaF\slash ITO\slash MeO-2PACz\slash Cs$_{0.05}$(FA$_{0.83}$MA$_{0.17}$)$_{0.95}$\allowbreak Pb(I$_{0.83}$Br$_{0.17}$)$_3$\slash C$_{60}$\slash BCP\slash Cu &
1.13 &
23.10 &
76.0 &
19.70 &
5.3 &
\cite{Tockhorn2025ACP} \\

Triple micro\slash nano structures &
glass\slash ARC\slash FTO\slash SnO$_2$ (IOE)\slash FA$_{0.65}$MA$_{0.35}$PbI$_{2.14}$Cl$_{0.86}$ (GPVK)\slash Spiro-OMeTAD\slash Au &
1.18 &
26.40 &
79.0 &
24.80 &
5.5 &
\cite{Cao2025CEJ} \\

Corrugated ETL &
ITO\slash PTAA\slash perovskite\slash PFBO-C12\slash BCP\slash Cu &
1.09 &
24.97 &
79.0 &
21.88 &
10.9 &
\cite{Liu2025OptLett} \\

Nanoimprinted grating on perovskite surface &
FTO\slash SnO$_2$\slash diphosphate\slash perovskite\slash MeO-PEAI\slash Spiro-OMeTAD\slash Ag &
1.17 &
25.38 &
82.9 &
24.63 &
9.7 &
\cite{Zhang2024Small} \\

PbI$_2$ inverse-opal scaffold\slash mesostructured precursor skeleton &
FTO\slash SnO$_2$\slash KCl\slash PbI$_2$ IO-derived perovskite\slash PEAI\slash Spiro-OMeTAD\slash Au &
1.17 &
25.23 &
84.43 &
24.93 &
13.2 &
\cite{Chen2025CEJ} \\

Holographic antireflection moth-eye film &
AR film\slash ITO\slash PTAA\slash MAPbI$_3$\slash C$_{60}$\slash BCP\slash Cu &
1.09 &
23.20 &
79.77 &
20.27 &
4.8 &
\cite{Wang2021OptLett} \\

Hyperuniform nanohole patterning in front glass &
Glass\slash ITO\slash TiO$_2$\slash MAPbI$_3$\slash Spiro-OMeTAD\slash Au &
1.13 &
23.92 &
87.66 &
23.62 &
12.3 &
This Work \\

\bottomrule
\end{tabularx}
\end{table*}

Secondly, we also compared the origin and nature of photonic enhancement.
Wang \textit{et al.} \cite{Wang2021OptLett} showed that the periodic grating-based antireflection film mainly works by suppressing Fresnel reflection and enhancing light capture at oblique angles. Tockhorn \textit{et al.} \cite{Tockhorn2025ACP} demonstrated that shallow periodic nanotextures reduced broadband reflection losses and raised the EQE-integrated photocurrent, while preserving absorber quality. By contrast, the present hyperuniform nanohole texture did not rely on a small set of discrete periodic diffraction channels. Instead, the disordered-but-correlated nanohole arrangement redistributed the incident field into a broader set of in-plane momentum components, which is advantageous for broadband coupling and for suppressing strong resonance-selective oscillations. 
From a photonic perspective, this is the key distinction of the present design: instead of maximizing performance through a single optimized grating period, we used statistical short-range order to achieve strong absorption enhancement over a broader spectral range, especially in the weakly absorbed long-wavelength region.

The contrast becomes even clearer when our work is compared with studies in which the nanostructure was introduced directly into transport layers, precursor scaffolds, or the perovskite itself. Liu \textit{et al.} \cite{Liu2025OptLett} formed a corrugated PFBO-C12 electron-transport layer with a 1.5~$\mu$m grating and achieved $\mathrm{J_{sc}=24.97}$~mAcm$^{-2}$ and $\mathrm{PCE=21.88}\%$, a 10.9\% PCE increase.
Zhang \textit{et al.} \cite{Zhang2024Small}  directly imprinted a grating into an ultrathin perovskite configuration and reported $\mathrm{J_{sc}=25.38}$~mAcm$^{-2}$, and $\mathrm{PCE=24.63\%}$, corresponding to a $10.9\%$ PCE increase. 
Chen \textit{et al.} \cite{Chen2025CEJ} employed a PbI$_2$ inverse-opal scaffold to control precursor infiltration and crystallization, reaching $\mathrm{V_{oc}=1.17}$~V, $\mathrm{J_{sc}=25.23}$~mAcm$^{-2}$, $\mathrm{FF=84.43}\%$, and $\mathrm{PCE=24.93}\%$, which is the largest PCE improvement ($13.2\%$) of \autoref{tab:nanotexture_comparison}. Cao \textit{et al.} \cite{Cao2025CEJ} integrated three optical micro/nano structures simultaneously, namely an ARC layer, an inverse-opal ETL, and a grating-perovskite layer, and reported $\mathrm{V_{oc}=1.18}$~V, $\mathrm{J_{sc}=26.40}$~mAcm$^{-2}$, $\mathrm{FF=79.0}\%$, and $\mathrm{PCE=24.80}\%$. These are all strong results, but in each case, the nanostructure directly altered electronically active regions or multiple internal interfaces, so the measured efficiency gain cannot be attributed solely to an optical effect.

From a fabrication standpoint, this distinction is also important. 
The grating-perovskite and ultrathin nanoimprinted approaches in the studies above required direct imprinting or structural transfer onto the perovskite layer, which intentionally altered crystallization behavior. 
The corrugated ETL approach demonstrated by Liu \textit{et al.} modified the electron-selective side of the junction, while the PbI$_2$ inverse-opal scaffold by Chen \textit{et al.} restructured the precursor-to-perovskite conversion pathway itself. By contrast, our proposed architecture confined the nanostructure to the front glass, thereby keeping the TiO$_2$/MAPbI$_3$/Spiro-OMeTAD junction planar. This design choice is attractive because it decouples light trapping from junction formation and avoids imposing nanoscale topography directly onto the absorber or transport layers. Although the aforementioned experimental studies achieved slightly higher champion PCE values, they did so by employing more invasive structural modifications to the active stack. The present design, therefore, offered a different trade-off: a comparatively large efficiency gain, together with the highest fill factor in \autoref{tab:nanotexture_comparison}, achieved without restructuring the photovoltaic junction itself.

Overall, the comparison in \autoref{tab:nanotexture_comparison} indicates that the main advantage of the present hyperuniform approach is not that it yields the highest absolute champion efficiency among all reported nanotextured PSCs, but that it achieves a high final efficiency of $23.62\%$ through a junction-preserving and predominantly optical strategy. 
Relative to the more passive front-side approaches in \autoref{tab:nanotexture_comparison}, it delivered a much larger efficiency gain. Compared with the abovementioned nanostructuring strategies, it maintained a distinctly higher fill factor and avoided direct modification of the electronically active interfaces. For this reason, hyperuniform front-glass patterning should be viewed as a particularly promising route for broadband light trapping when the goal is to improve photocurrent while maintaining the simplicity and electrical integrity of a planar perovskite junction.

\section{Conclusion}

In this work, we proposed front-glass hyperuniform nanohole patterning as an effective junction-preserving light-management strategy for planar MAPbI$_3$ perovskite solar cells. By introducing the nature-inspired hyperuniform texture into the top glass rather than the electronically active layers, we enhanced optical coupling and extended the effective photon path length while preserving the underlying planar heterojunction architecture. Three-dimensional optical simulations revealed that the hyperuniform nanohole layer redistributed incident light over a broader range of in-plane momentum states, strengthened near-interface electromagnetic fields, and improved long-wavelength coupling into the absorber, thereby promoting broadband absorption enhancement beyond that of planar and periodic reference structures. The texture also exhibited weak polarization dependence, strong tolerance to oblique illumination, and reduced sensitivity to thickness-dependent interference effects, highlighting the robustness of the proposed optical design. By coupling the optical generation profiles with drift-diffusion electrical modeling, we further demonstrated that these photonic enhancements directly translate into improved device performance.
The optimized hyperuniform architecture resulted in an improved PCE of 23.62\%, largely due to an increase in $\mathrm{J_{sc}}$ from 21.57 to 23.92 mA/cm$^2$ after pattern integration. 
Other key metrics remained unaffected, with the $\mathrm{V_{oc}}$ and fill factor holding steady at 1.13 V and 87.66\%, respectively.
The optimized hyperuniform architecture achieved an improved PCE of 23.62\%, driven primarily by the increase in $\mathrm{J_{sc}}$ from 21.57 to 23.92 mAcm$^{-2}$ following pattern integration, while $\mathrm{V_{oc}}$ and the fill factor remained almost unchanged at 1.13 V and 87.66\%, respectively. 
Systematic benchmarking against planar and periodic nanohole counterparts confirmed that the performance gain arises not only from surface texturing but also from the distinct optical response enabled by hyperuniform spatial statistics. 
Optical analysis across multiple realizations of hyperuniform configurations revealed only marginal variation in performance.
To examine fabrication-relevant dimensional perturbations, we further conducted a stochastic radius-disorder analysis on numerous hyperuniform realizations. 
The observed photocurrent enhancement remained predominantly intact under these variations, supporting the proposed architecture as a fabrication-tolerant and robust strategy for photovoltaic enhancement.
Taken together, these results demonstrate that hyperuniform front-substrate nanopatterning provides a promising and physically realistic route toward broadband, angle-resilient, and electrically compatible light trapping in high-performance perovskite photovoltaics.

\begin{acknowledgement}


A.S. acknowledges financial support from the Bangladesh University of Engineering and Technology (BUET) through its Postgraduate Fellowship Program. The authors express their gratitude to the Department of Electrical and
Electronic Engineering at Bangladesh University of Engineering and Technology (BUET) for providing access to the Ansys Lumerical software and the necessary technical support, including computation facilities.

\end{acknowledgement}

\begin{datastatement}
The data that support the findings of this study are available from the corresponding author upon reasonable request.
\end{datastatement}

\begin{suppinfo}

Complex refractive index of different layers in PSC, Electrical simulation parameters for various layers in PSC, Thickness and doping optimizations of ETL and HTL layers, Hyperuniform pattern generation algorithm, and Fabrication process of hyperuniform patterning in glass layers.

\end{suppinfo}

\begin{conflictstatement}
A preprint version of this work was previously made available: Sur, A.; Nath, K.; Zubair, A. Nature-Inspired Hyperuniform Nanohole Patterning for Robust Broadband Absorption Enhancement in Perovskite Solar Cells. 2026, arXiv preprint arXiv:2604.11264. 10.48550/ arXiv.2604.11264 (accessed April 16, 2026). The authors declare no competing financial interest.
\end{conflictstatement}

\bibliography{mybib}

\end{document}